\documentclass[12pt,a4paper,twoside]{article}

\usepackage{amsmath,amssymb,amsthm}
\makeatletter
\renewcommand{\@seccntformat}[1]{\csname the#1\endcsname.\hspace{0.5em}}
\makeatother

\newcommand{\bA}{{\bf A}}
\newcommand{\ba}{{\bf a}}
\newcommand{\br}{{\bf r}}
\newcommand{\BB}{{\bf B}}
\newcommand{\bfe}{{\bf e}}

\newcommand{\C}{\mathbb{C}}
\newcommand{\R}{\mathbb{R}}
\newcommand{\Z}{\mathbb{Z}}
\newcommand{\N}{\mathbb{N}}

\renewcommand{\Im}{\operatorname{Im}}
\renewcommand{\Re}{\operatorname{Re}}
\newcommand{\sgrad}{\operatorname{sgrad}}
\newcommand{\grad}{\operatorname{grad}}
\renewcommand{\ln}{\operatorname{ln}}
\newcommand{\sh}{\operatorname{sh}}
\newcommand{\ch}{\operatorname{ch}}
\newcommand{\supp}{\operatorname{supp}}
\newcommand{\ctg}{\operatorname{ctg}}

\newcommand{\be}{\begin{equation}}
\newcommand{\ee}{\end{equation}}
\newcommand{\dd}{\partial}

\renewcommand{\phi}{\varphi}
\renewcommand{\epsilon}{\varepsilon}
\newcommand{\bra}{\langle\,}
\newcommand{\ket}{\,\rangle}

\newtheorem{thm}{Theorem}[section]
\newtheorem{lem}[thm]{Lemma}
\newtheorem{prop}[thm]{Proposition}
\newtheorem{cor}[thm]{Corollary}
\theoremstyle{remark}
\makeatletter
\renewcommand{\th@remark}{%
  \thm@headfont{\normalfont\bfseries}%
  \normalfont 
  \thm@preskip\topsep \divide\thm@preskip\tw@
  \thm@postskip\thm@preskip
}
\makeatother
\newtheorem{rem}[thm]{Remark}
\newtheorem*{rem*}{Remark}

\begin{document}

\title{Zero modes in a system of Aharonov--Bohm fluxes}

\author{V.~A. Geyler\\
\emph{\normalsize Department of Mathematics, Mordovian State University}\\
\emph{\normalsize Bolshevistskaya 68, Saransk 430000, Russia}\\
\\
P. \v{S}\v{t}ov\'{\i}\v{c}ek\\
\emph{\normalsize Department of Mathematics, Faculty of Nuclear Science
}\\
\emph{\normalsize Czech Technical University}\\
\emph{\normalsize Trojanova 13, 120 00 Prague, Czech Republic}}

\date{{}}
\maketitle
\begin{abstract}
  \noindent We study zero modes of two-dimensional Pauli operators
  with Aharonov--Bohm fluxes in the case when the solenoids are
  arranged in periodic structures like chains or lattices. We also
  consider perturbations to such periodic systems which may be
  infinite and irregular but they are always supposed to be
  sufficiently scarce.
\end{abstract}

\section{Introduction}

The appearance of zero modes (wave functions at zero energy which are
ground states for a positive quantum Hamiltonian) belongs to the most
interesting phenomena in systems with topologically non-trivial
configuration spaces; see the discussion and an extensive bibliography
in \cite{BRFHMM}. Zero modes of the Dirac and Pauli operators are of
great importance in many places in quantum field theory and
mathematical physics \cite{ASi,JR1,JR2}. They are the ingredients for
the computation of the index of these operators and play a key role in
understanding anomalies. One of the best known examples for such
operators is the Pauli Hamiltonian of a two-dimensional charged
particle moving in a magnetic field perpendicular to a plane and
penetrating the plane in a bounded domain. In this case the field
defines a vector bundle with a non-trivial connection and zero modes
appear at sufficiently high strength of the field \cite{AC}. More
precisely, it is easy to prove that the dimension $d$ of the space of
zero modes is $d=\lfloor|\Phi|\rfloor$ where $\Phi$ is the total flux
of the magnetic field measured in magnetic flux quanta, and for a real
$x$, $x\ge0$, $\lfloor{}x\rfloor$ denotes the lower integer part of
$x$ ($\lfloor0\rfloor=0$, $\lfloor n\rfloor=n-1$ for $n\ge1$ integer,
and otherwise $\lfloor x\rfloor=[x]$, the integer part of $x$). It is
worthy to note that in the three-dimensional case the appearance and
the degeneracy of zero modes is a more subtle fact (see e.g.
\cite{AMN1,AMN2,AMN3,AMN5,AMN6,ES} and the discussion therein).

In the current paper we restrict our consideration to two-dimensional
systems only. More precisely, we consider Pauli operators which are
Hamiltonians of an electron confined to a plane and subjected to a
perpendicular time-independent magnetic field which is the sum of a
uniform field and an additional field contributed by a (finite or
infinite) array of singular flux tubes or, in other words, by an array
of solenoids of zero width. We focus on zero modes in such systems. In
more detail, the aim of the paper is to find conditions for appearance
of zero modes in systems placed in a magnetic field with an infinite
array of Aharonov--Bohm vortices. It has been shown in \cite{GG} on
the physical level of rigor that zero modes occur if Aharonov--Bohm
vortices are arranged in a periodic plane lattice provided that not
all magnetic fluxes involved have integer values. In this paper we
present a rigorous proof and show that under the same condition
imposed on the flux, the result is true for a chain of Aharonov--Bohm
solenoids or, more generally, for a uniformly discrete union of such
chains. Moreover, the zero modes are retained if one adds to such a
periodic structure of Aharonov--Bohm solenoids a not necessarily
regular array of solenoids having sufficiently low density. This
stability of zero-modes for the Hamiltonian that we call
$H_{\rm{max}}$ (its definition is discussed in
Section~\ref{sec:some-comm-hist}) shows that their origin differs from
that for localized states in the so called Aharonov--Bohm cages
\cite{VMD,VBDM}, the latter are destroyed by arbitrarily small period
modulations \cite{Oh}.

The main results of the paper are obtained with the help of a version
of the Aharonov--Casher ansatz \cite{AC}. This version was proposed by
Dubrovin and Novikov in \cite{DN} who employed it for an explicit
construction of ground states of periodic magnetic Schr\"odinger
operators (see Novikov's review paper \cite{Nov}). In our case, this
ansatz reduces the problem of finding zero-modes to some estimates for
entire functions. The mechanism of appearance of zero modes in the
considered cases is close to that for a two-dimensional system in a
uniform magnetic field in the presence of an infinite array of point
scatterers \cite{Gey,ARB,AR,AAG,GM,GZAA}.

An interesting physical consequence of our result is the occurrence of
oscillations of the type ``localization--delocalization'' in periodic
systems of Aharonov--Bohm solenoids placed in a varying uniform
magnetic field (Theorem \ref{orig:thm-6.8}). Another interesting
result described in Theorems \ref{orig:thm-6.5} and \ref{orig:thm-6.8}
is related to the problem of absolute continuity of the spectrum of
the Schr\"odinger operator with periodic {\it vector } potential
${\bf{A}}$. This absolute continuity has been proved for a wide class
of potentials ${\bf{A}}$ \cite{BS1,BS2,BSS,KL,Mor,Sob}. An example of
a vector potential ${\bf{A}}$ having eigenvalues in the spectrum of
the corresponding Schr\"odinger operator was given in \cite{Fil} but
only for dimensions higher than 3. Our results give such an example in
dimension 2.

The paper is organized as follows. In Section~\ref{sec:some-comm-hist}
we try to point out some aspects regarding the history and the
background of the problem. In Section~\ref{sec:pauliop-orig:1} we
discuss shortly the gauge invariance in the case when the magnetic
field is a distribution. In Section~\ref{sec:examples-orig:2} we
introduce several basic examples of models with Aharonov--Bohm fluxes
some of them are the main subject of this paper and are studied in
detail in the sequel. Section~\ref{sec:regdef-orig:3} is devoted to a
rigorous definition of the Pauli operator with Aharonov--Bohm fluxes.
In Section~\ref{sec:intflux-orig:4} we discuss the elimination of
integer-valued Aharonov--Bohm fluxes. In
Section~\ref{sec:groundsts-orig:5} we recall the Aharonov--Casher
ansatz which makes it possible to construct ground states of the Pauli
operator using the theory of analytic functions. The main results of
the paper are contained in Sections~\ref{sec:zeromod-orig:6} and
\ref{sec:conserv-orig:7}. In Section~\ref{sec:zeromod-orig:6} we study
zero modes of the Pauli operator with an infinite periodic system of
Aharonov--Bohm solenoids. In Section~\ref{sec:conserv-orig:7} we
address the question of perturbations of such periodic structures
caused by translations and additions of Aharonov--Bohm solenoids. The
subsystem formed by solenoids affected by the perturbation may be
infinite and irregular but we always suppose that it is sufficiently
scarce. Here we also discuss some examples of irregular Aharonov--Bohm
systems. For the reader's convenience we have included three
appendices. In the first appendix we collect some basic definitions
and auxiliary results concerning lattices.  In the second appendix we
recall some basic notions and results from the theory of analytic
functions related to the growth of entire functions. The third
appendix is devoted to the Weierstrass $\sigma$-function.

\section{Additional comments on the history and the background of the problem}
\label{sec:some-comm-hist}

There are many interesting and important physical problems related to
systems involving Aharonov--Bohm fluxes. Since the publication of the
original paper due to Aharonov and Bohm \cite{AB} the physics of a
magnetic flux in an infinitely thin solenoid (called {\it
  Aharonov--Bohm flux} or {\it Aharonov--Bohm vortex}) has been
investigated both from theoretical and experimental points of view
\cite{OP,Ham}. The physical origin of the Aharonov--Bohm effect is
even a subject of theoretical investigations up to now \cite{AK}.  On
the other hand, the motion of a charged particle (an electron, a hole
or a composite fermion) in a plane perpendicular to a uniform magnetic
field has found an important application in physics of the quantum
Hall effect \cite{KDP,QHE}. The most striking feature of the
Hamiltonian of such a system is the Landau quantization of the
spectrum which consists of highly degenerated equidistant energy
levels; this makes quantum Hall phenomena possible. Moreover, it is of
interest to know how the quantum Hall system is altered by various
defects, in particular, by impurities or by inhomogeneities of the
magnetic field.  Additional Aharonov--Bohm fluxes appear to be a
minimal modification of the uniform magnetic field, while general
inhomogeneous magnetic fields are extremely difficult to handle
\cite{Thi,Thi1}. Similarly, a minimal perturbation of the quantum Hall
system is given by a point perturbation of the Landau operator (i.e.,
the Schr\"odinger operator with a uniform magnetic field) \cite{Gey}.
As shown below, both modifications require the operator extension
theory for a correct construction of the corresponding Hamiltonian
\cite{AGHH}.

The vector potential of a system of Aharonov--Bohm solenoids has a
strong singularity at the points where the plane intersects the
solenoids. Therefore the differential operator defining the
Hamiltonian is not essentially self-adjoint on its natural domain.
This is true both in the non-relativistic case (for the Schr\"odinger
and Pauli operators) and the relativistic one (for the Dirac
operator). The boundary conditions for Schr\"odinger operators with an
Aharonov--Bohm vortex as well as the corresponding self-adjoint
extensions (i.e., Hamiltonians describing a spinless non-relativistic
quantum particle) are considered in many papers, let us mention e.g.
\cite{Rui,AJS,AT,DSt}. The multi-solenoid case is more difficult
because of the rotational symmetry violation. This case was treated by
means of the Krein resolvent formula in \cite{Sto1}, and for an
infinite chain of solenoids in \cite{Sto2}; different approaches are
presented in \cite{Nam,IT1,IT2}. The problem of defining the boundary
conditions at the presence of a uniform background field has been
investigated in \cite{ESV,Min}. In the relativistic case, the problem
of defining the appropriate Dirac operator is discussed e.g. in
\cite{dSG,Hag1,BDS}, and at the presence of a uniform component -- in
the recent articles \cite{Ogu,GGS,GGS1,GGS2}. In all the mentioned
papers, the spectral or scattering properties of the derived
Hamiltonians are studied as well.

On sufficiently smooth functions from
$L^2(\R^2)\otimes\C^2=L^2(\R^2;\C^2)$ the two-dimensional Pauli
operator for a charged particle with the spin $s$ and the gyromagnetic
ratio $g$ acts as a formal differential operator \cite{LL}
\begin{equation}
  \label{1.1}
  \hat H\equiv \hat H(\bA)
  =\frac{1}{2m^*}\left(\frac{\hbar}{i}\nabla-\frac{e}{c}\bA\right)^2
  -\hat\mu\BB
\end{equation}
where $e$ and $m^*$ are the charge and the mass of the particle,
respectively, $\bA=(A_x,A_y)$ is the vector potential of a magnetic
field $\BB=B\bfe_z$, $B=\dd_xA_y-\dd_yA_x$, $\hat\mu$ is the magnetic
momentum operator,
$$
\hat\mu=gs\mu_B\hat s_z\,,
$$
with $\mu_B$ being the Bohr magneton, $\mu_B=-|e|\hbar/(2m^*c)$,
and
$$
\hat s_z =\frac{1}{2}\, \sigma_z=\frac{1}{2}\left(
  \begin{array}{cc}
    1&0\\
    0&-1
  \end{array}
\right)\,
$$
(we consider the motion of a particle in the plane $\R^2$
canonically embedded in the space $\R^3$). In general, the
non-relativistic limit of the Dirac equation leads to the value $g=2$,
and the main part of our work deals with this value of the
gyromagnetic ratio. In the case of an Aharonov--Bohm solenoid $\BB$ is
proportional to the Dirac delta function, $\delta(\br)$, and therefore
the operator (\ref{1.1}) takes the form of the Schr\"odinger operator
perturbed at a point and with a finite coupling constant $\alpha$
standing in front of the ``$\delta$-potential''. On the other hand, it
is well known that in the two-dimensional case under consideration the
expression (\ref{1.1}) defines a non-trivial perturbation of the
operator
\begin{equation}
  \label{1.1a}
  \hat H_0\equiv \hat H_0(\bA)
  =\frac{1}{2m^*}\left(\frac{\hbar}{i}\nabla-\frac{e}{c}\bA\right)^2
\end{equation}
only if $\alpha$ is in some sense "infinitesimal" \cite{AGHH} (we
suppose that appropriate boundary conditions defining $\hat H_0$ are
chosen). This problem has been analyzed in \cite{VB} in detail for an
arbitrary positive value of $g$. To get around it, a solenoid of
finite radius $R$ is considered with a rotationally symmetric magnetic
flux inside the solenoid but otherwise having an arbitrary profile
(including the magnetic flux supported on the surface of the infinite
cylinder), and the limit $R\to 0$ is discussed. In addition to
\cite{VB} let us also mention papers
\cite{BV,CNFT,CdC,CFdC,Hag3,Par,PO,PYo,PY}. Of course, the same
approach is useful when a uniform component of the field is present or
in the case of the Dirac operator (see \cite{EHO,Tam,Tam1} and
references therein).

In the most important case when the gyromagnetic ratio $g$ equals $2$
the Pauli operator has remarkable supersymmetry properties which makes
it possible to use the Aharonov--Casher decomposition \cite{AC}.  As a
result, we have a convenient definition of the Pauli operator with a
singular potential by means of a quadratic form (see
Section~\ref{sec:regdef-orig:3}). More precisely, in this case we
have, as usual, two natural quadratic forms associated to the
expression (\ref{1.1}) -- the minimal and the maximal one (with the
definition of the magnetic Schr\"odinger operator taken from
\cite{CFKS}). These forms provide us with two natural types of
self-adjoint operators denoted $H^{\pm}_{\max}$ and $H^{\pm}_{\min}$
and playing the role of Pauli operators with Aharonov--Bohm solenoids
(the sign $\pm$ stands for spin up and spin down supersymmetric
partners, respectively). There is an important distinction between the
operators $H_{\rm min}$ and $H_{\rm max}$.  As it follows from the
definitions, both operators $H^\pm_{\rm min}$ coincide with the
Friedrichs extension of the symmetric operator defined by expression
(\ref{1.1a}) with the vector potential $\bA$ corresponding to a system
of Aharonov--Bohm fluxes. Therefore this extension (denoted simply by
$H_{\rm min}$) may be interpreted as the Hamiltonian of a ``spinless''
particle moving in a system of Aharonov--Bohm fluxes (this corresponds
to physical problems for an electron when the spin--orbit coupling can
be neglected and spin splitting is taken into account with the help of
the perturbation theory \cite{LL}). Such a Hamiltonian has been
considered e.g., in \cite{AJS,VB}. On the other hand, the operators
$H^{\pm}_{\rm max}$ do not coincide in general which indicates that
they directly take into account the energy of the spin--orbit
interaction and therefore they may be regarded as the Pauli operators
of the system under consideration. In the present article we
concentrate mainly on zero modes of $H_{\max}$. Note that boundary
conditions defining the Hamiltonian $H_{\max}$ are given in
\cite{AT,DSt} (in the case of a single solenoid) and in \cite{GS} (in
the two-solenoid case).

For a finite system of Aharonov--Bohm solenoids, the existence problem
of zero-energy eigenfunctions was considered in
\cite{Ara1,Ara2,Ara3,Ara4,Ara5}. In this case the number $d$ of
linearly independent zero-modes depends on the fractional parts of
fluxes in the individual solenoids, $\{x\}=x-[x]$, rather than only on
the total flux $\Phi$ in the system. This phenomenon is a consequence
of the gauge invariance properties for the Aharonov--Bohm fluxes (see
e.g. papers \cite{Hel,HHN,HHHO,HHHO1}).  In the case when the
considered magnetic field has a ``regular'' component in addition to
the magnetic field of Aharonov--Bohm solenoids the appearance of zero
modes has been analyzed in \cite{HO,EV}. The results of \cite{EV} are
applicable also to the case when an infinite number of Aharonov--Bohm
solenoids is present in the system but the total magnetic flux is
necessarily {\it finite} (moreover, after some gauge transformation
the total variation of the flux must be finite). On the other hand, it
is clear that the thermodynamic limit of a bounded system with a fixed
density of Aharonov--Bohm fluxes is a system with an infinite number
of Aharonov--Bohm solenoids and with an {\it infinite} total flux. An
example for a system of such a kind is the quasi-two-dimensional
system with columnar defects in a uniform magnetic field directed
along the defect axis \cite{LLMK,LLFM,Lat} or the GaAs/AlGaAs
heterostructure coated with a film of type-II superconductor
\cite{BKP} (in the latter case the Aharonov--Bohm fluxes are arranged
in a honeycomb lattice, the so-called Abrikosov lattice).

As for the spectral properties of the operator $H_{\min}$, they
have been investigated recently in detail by Melgaard, Ouhabaz and
Rozenblum \cite{MOR}. In particular, these authors proved with the
help of results from \cite{Bal} and \cite{LW} that $H_{\min}$ has
no zero modes at least for periodic lattices of Aharonov--Bohm
solenoids, and therefore it differs from $H^+_{\max}$ and
$H^-_{\max}$ for generic values of magnetic fluxes (and even it is
not unitarily equivalent to these operators). Let us note that it
is possible to extend this result to a chain of Aharonov--Bohm
solenoids.

\section{The Pauli operator with a singular magnetic field}
\label{sec:pauliop-orig:1}

In what follows we consider the motion of an electron with the
gyromagnetic ratio $g=2$, therefore
\begin{equation}
  \label{1.2}
  \hat H=\frac{\hbar^2}{2m^*}\left[\left(i\dd_x+\frac{e}{c\hbar}A_x\right)^2+
    \left(i\dd_y+\frac{e}{c\hbar}A_y\right)^2
    -\frac{e}{c\hbar}\sigma_zB\right]\,.
\end{equation}
Let us denote for simplicity
\begin{equation}
  \label{1.3}
  \frac{e}{c\hbar}\bA=\ba\,,\quad
  \frac{e}{c\hbar}B=b\,,
\end{equation}
so that $\dd_xa_y-\dd_ya_x=b$. In order to employ the dimensionless
units we shall consider the operator
\begin{equation}
                   \label{1.3a}
H\equiv H(\ba)=\frac{2m^*}{\hbar^2}\hat H(\bA)\,.
\end{equation}
Introducing a quantum of the magnetic flux,
\begin{equation}
  \label{1.4}
  \Phi^0=\frac{2\pi c\hbar}{e}\,,
\end{equation}
we also have
\begin{equation}
  \label{1.5}
  \ba=\frac{2\pi}{\Phi^0}\bA,\quad b=\frac{2\pi}{\Phi^0}B\,,
\end{equation}
\begin{equation}
  \label{1.6}
  H\equiv H(\ba)=(i\dd_x+a_x)^2+(i\dd_y+a_y)^2-\sigma_z b\,.
\end{equation}
The operator $H$ (and respectively the operator $\hat H$) decomposes
in a sum of two scalar operators,
\begin{equation}
  \label{1.6a}
  H^{\pm}\equiv H^{\pm}(\ba)=(i\dd_x+a_x)^2+(i\dd_y+a_y)^2\mp b\,,
\end{equation}
(respectively $\hat H^{\pm}(\bA)\equiv \hat H^{\pm}$) acting in
$L^2(\R^2)$. We admit the vector potential $\ba$ to have singular
points, more precisely, we assume that
\begin{equation}
  \label{1.7}
  a_x\,,\,\,a_y\in L^1_{\rm loc}(\R^2)\cap
  C^{\infty}(\R^2\setminus\Omega)
\end{equation}
where $\Omega$ is a discrete subset (possibly finite or empty) in
$\R^2$. Consequently, the magnetic field $b=\dd_xa_y-\dd_ya_x$ is,
in general, a distribution in $\R^2$ whose singular support is
contained in $\Omega$. Expressions (\ref{1.1}) and (\ref{1.6a})
represent symmetric operators with the domain
$C^{\infty}_0(\R^2\setminus\Omega)$; these operators will be
denoted $\hat H^{\pm}(\bA,\Omega)$ and $H^{\pm}(\ba,\Omega)$,
respectively. If the singular support of $B$ coincides with
$\Omega$ (in this case $\Omega$ is determined by the vector
potential $\bA$) we shall simply write $\hat H^{\pm}(\bA)$ and
$H^{\pm}(\ba)$.

It is important to note that also in the case when $b$ is a
distribution the operator $H^{\pm}(\ba)$ depends, up to unitary
equivalence, only on $b$. More precisely, we have the following
proposition.

\begin{thm}[gauge invariance of the operator $H^{\pm}(\ba)$]
  \label{orig:thm-1.1}
  Let $\ba$ and $\tilde\ba$ be vector potentials with the same
  magnetic field $b$ (i.e,
  $\ba,\,\tilde\ba\in{}L^1_{\rm{loc}}(\R^2;\R^2)\cap{}C^{\infty}
  (\R^2\setminus\Omega;\R^2)$ and
  $\dd_xa_y-\dd_ya_x=\dd_x\tilde{a}_y-\dd_y\tilde{a}_x=b$ in the sense
  of distributions). Then the operators $H^{\pm}(\ba,\Omega)$ and
  $H^{\pm}(\tilde\ba,\Omega)$ are unitarily equivalent. In more
  detail, there exists a~real-valued function $f$ belonging to
  $C^{\infty}(\R^2\setminus\Omega)$ such that
  $\tilde\ba=\ba+\grad{}f$, and
  $H^{\pm}(\tilde\ba,\Omega)=W^{-1}H^{\pm}(\ba,\Omega)W$ where $W$ is
  the unitary operator acting via multiplication by the function
  $\exp(-if)$.
\end{thm}

Of course, this theorem is well known in the case when the field $b$
is a function (not a distribution). In the case when $b$ is a
distribution the theorem is a consequence of the following lemma whose
elementary proof was communicated to us by K.~V. Pankrashkin.

\begin{lem}\label{orig:lem-1.1}
  Assume that
  $\ba\in{}L^1_{\rm{loc}}(\R^2)\cap{}C^{\infty}(\R^2\setminus\Omega)$
  and the equality $\dd_xa_y-\dd_ya_x=0$ holds true in $\R^2$ in the
  sense of distributions. Let $\omega\in\Omega$ and let $Q$ be a
  rectangle containing $\omega$ but no other points from $\Omega$.
  Then
  \begin{equation}
    \label{1.7a}
    \int\limits_{\dd Q}\,a_x\,dx+a_y\,dy=0\,.
  \end{equation}
\end{lem}

\begin{proof}
  Let us choose functions $\phi,\,\psi\in C_0^{\infty}(\R^2)$ so that
  $\omega\notin\supp\phi$, $\phi(x,y)=1$ in some neighborhood of the
  boundary $\dd Q$, $\psi(x,y)=\phi(x,y)$ on $\R^2\setminus Q$ and
  $\psi(x,y)=1$ on $Q$. Using the Green formula we obtain
  \begin{eqnarray*}
    && \int\limits_{\dd Q}\,a_x\,dx+a_y\,dy
    = \int\limits_{\dd Q}\,\phi a_x\,dx+\phi a_y\,dy
    = \iint\limits_{Q}\,
    \left(\dd_x(\phi a_y)-\dd_y(\phi a_x)\right)\,dxdy \\
    &&=\, \iint\limits_{Q\cap\,\supp(\varphi)}\,
    \phi\left(\dd_x a_y-\dd_ya_x\right)\,dxdy+
    \iint\limits_{Q}\,
    \left( a_y\dd_x\phi- a_x\dd_y\phi\right)\,dxdy \\
    &&=\, \iint\limits_{\R^2}\,
    \left( a_y\dd_x(\phi-\psi)- a_x\dd_y(\phi-\psi)\right)\,dxdy
    =0\,.
  \end{eqnarray*}
  Here we have used the fact that the expression $\dd_xa_y-\dd_ya_x$
  represents a smooth function on $\R^2\setminus\Omega$ which
  necessarily vanishes on this domain.
\end{proof}

\begin{proof}[Proof of Theorem~\ref{orig:thm-1.1}]
  From Lemma~\ref{orig:lem-1.1} we derive in a standard manner that if
  $\dd_xa_y-\dd_ya_x=0$ on $\R^2$ in the sense of distributions then
  there exists a~real-valued function
  $f\in{}C^{\infty}(\R^2\setminus\Omega)$ such that $\ba=\grad{}f$ on
  $\R^2\setminus\Omega$. Consequently, if $\ba$ and $\tilde\ba$ obey
  the assumptions of the theorem then for some function $f\in
  C^{\infty}(\R^2\setminus\Omega)$ we have $\tilde\ba=\ba+\grad{}f$.
  Let us denote by $W$ the operator acting via multiplication by the
  function $\exp(-if)$. Clearly, $W$ is a well defined unitary
  operator in $L^2(\R^2)$. Moreover, $W$ leaves invariant the subspace
  $C_0^{\infty}(\R^2\setminus\Omega)$. A simple computation shows that
  $W^{-1}H^{\pm}(\ba,\Omega)W=H^{\pm}(\tilde\ba,\Omega)$.  Hence the
  operators $H^{\pm}(\ba,\Omega)$ and $H^{\pm}(\tilde\ba,\Omega)$ are
  unitarily equivalent.
\end{proof}

\begin{rem}
  Clearly, if $\ba=\grad{}f$ in the sense of distributions then
  $\dd_xa_y-\dd_ya_x=0$ in the same sense.
\end{rem}

\begin{rem}
  A proposition analogous to that of Theorem~\ref{orig:thm-1.1} is
  also valid for the operator $\hat{H}^{\pm}(\bA,\Omega)$. Namely, if
  $\dd_xA_y-\dd_yA_x=\dd_y\tilde A_x-\dd_x\tilde A_y=B$ then
  $\tilde\bA=\bA+\grad{}f$ and
  $\hat{H}^{\pm}(\bA,\Omega)=W^{-1}\hat{H}^{\pm}(\tilde\bA,\Omega)W$
  where $W=\exp(-(ie/c\hbar)f)$.
\end{rem}

Owing to the gauge invariance it is possible to require the vector
potential $\bA$ to have some additional properties. For example, the
vector potential $\bA$ can be frequently chosen so that it fulfills
the Lorentz gauge condition
\begin{equation}
  \label{1.7b}
  {\rm div}\,\bA=0.
\end{equation}

\section{Basic examples}
\label{sec:examples-orig:2}

In this section we recall several basic examples of magnetic fields
fulfilling condition (\ref{1.7}). At the same time, we introduce the
necessary notation. The majority of results presented in the current
paper concern Examples~5, 6 and 7. In what follows it will be
convenient to identify the Euclidean plane $\R^2$ with the complex
plane $\C$ and to work with the complex coordinates $z=x+iy$ and $\bar
z=x-iy$.

\subsubsection*{Example~1. The homogeneous field}

In this case $B={\rm const}$ by definition and one can set
$$
A_x=-\frac{B}{2}\,y,\textrm{ } A_y=\frac{B}{2}\,x
$$
(the symmetric gauge). In the complex coordinates we have
$$
A_x=\frac{B}{2}\,\Im\bar z\,,
\textrm{ } A_y=\frac{B}{2}\,\Re\bar z\,.
$$
In this example $b=2\pi\xi$ where $\xi$ is the number of magnetic
flux quanta through a unit area in $\R^2$ (the flux density). The
Lorentz gauge condition (\ref{1.7b}) is obviously fulfilled.

\subsubsection*{Example~2. The magnetic field of an Aharonov--Bohm solenoid}

Here $B(\br)=\Phi\delta(\br)$ where $\Phi$ is the magnetic flux
through the solenoid. In this case one can set
$$
A_x=-\frac {\Phi}{2\pi}\,\frac{y}{r^2}\,,
\quad A_y=\frac {\Phi}{2\pi}\,\frac{x}{r^2}\,.
$$
Equivalently,
$$
a_x=\theta\,\Im\frac{1}{z}\,,
\quad a_y=\theta\,\Re\frac{1}{z}\,,
$$
where $\theta=\Phi/\Phi^0$ is the number of magnetic flux quanta
through the Aharonov--Bohm solenoid. Actually, it is well known that
$$
\Delta\,\ln(|z|)=2\pi\delta(z)\,.
$$
In the local coordinates we have
$$
B = \frac{\dd}{\dd x}A_y-\frac{\dd}{\dd y}A_x
= \frac{\Phi}{2\pi}\left(\frac{\dd}{\dd x}\Re\frac{1}{z}
- \frac{\dd}{\dd y}\Im\frac{1}{z}\right)\!
= \frac{\Phi}{2\pi}\left(\frac{\dd^2}{\dd x^2}
+ \frac{\dd^2}{\dd y^2}\right)\ln|z|
= \Phi\,\delta(z).
$$
The vector potential $\ba$ can be also written as
\begin{equation}
  \label{2.11a}
  \ba= \theta\, \sgrad\ln|z|\,.
\end{equation}
Here and everywhere in what follows $\sgrad$ stands for the symplectic
gradient,
\begin{equation}
  \label{2.12b}
  \sgrad f=\left(-\frac{\dd}{\dd y}f\,,\frac{\dd}{\dd x}f\right)\,.
\end{equation}
Hence $b=2\pi\theta\delta(z)$. The equality ${\rm div}\,\ba=0$
trivially follows from (\ref{2.11a}).

\subsubsection*{Example~3. An arbitrary system of Aharonov--Bohm solenoids}

Let now $\Omega$ be a discrete subset of the plane $\R^2$ and let
$(\Phi_\omega)_{\omega\in\Omega}$ be an arbitrary family of real
numbers with indices from $\Omega$. We shall consider a system of
Aharonov--Bohm fluxes intersecting the plane in the points from the
set $\Omega$ and perpendicular to the plane. The number $\Phi_\omega$
equals the flux in the solenoid passing through the point
$\omega\in\Omega$. Then
\begin{displaymath}
  b = \frac{2\pi}{\Phi^0}\,B
  = 2\pi\sum\limits_{\omega\in\Omega}\theta_\omega\delta(z-\omega)
\end{displaymath}
where, of course, $\theta_\omega=\Phi_\omega/\Phi^0$ is the number of
magnetic flux quanta through the solenoid $\omega$. For a vector
potential $\ba$ fulfilling the Landau gauge condition one can choose a
meromorphic function $M(z)$ with the following properties:

\medskip

\noindent 1) $M(z)$ has simple poles only,

\noindent 2) the set of poles of $M(z)$ coincides with $\Omega$,

\noindent 3) the residue of $M(z)$ at the point $\omega$
equals $\theta_\omega$.

\medskip

\noindent According to the Mittag-Leffler theorem such a
function always exists. 
The computations carried out in Example 2 (jointly with the
Cauchy--Riemann conditions) show that one can set
$$
a_x(z,\bar z)=\Im M(z)\,, \quad a_y(z,\bar z)=\Re  M(z)\,.
$$

The operator $H^{\pm}(\ba)$ will be also denoted by the symbol
$H^{\pm}(\Omega,\Theta)$ where
$\Theta=(\theta_\omega)_{\omega\in\Omega}$. The couple
$(\Omega,\Theta)$ determines the operator $H^{\pm}(\Omega,\,\Theta)$
unambiguously up to unitary equivalence.

\subsubsection*{Example~4. An arbitrary system of Aharonov--Bohm solenoids
  with fluxes taking a finite number of values}

Separately we consider the case when the number of mutually
different fluxes in the family $(\Phi_\omega)_{\omega\in\Omega}$
is finite (equivalently, the family
$(\theta_\omega)_{\omega\in\Omega}$ contains only a finite number
of mutually different numbers $\theta_\omega$). We start from the
case when all the involved solenoids carry the same flux:
$\theta_\omega=\theta$, $\forall \omega\in\Omega$. In this case we
always set
$$
M(z)=\theta\,\frac{W'(z)}{W(z)}\,.
$$
Here the function $W(z)$ differs from the Weierstrass canonical
product $W_\Omega(z)$ related to the set $\Omega$ only by a
multiplier $\exp(g(z))$ where $g(z)$ is an entire function. 
Obviously, the set of poles of the function $W'(z)/W(z)$ coincides
with $\Omega$, all the poles are simple and all the residues are
equal to 1. Thus one can set
\begin{equation}
  \label{s2.0}
  \ba=\theta\,{\sgrad\ln(|W(z)|)}\,.
\end{equation}
Actually, locally we have
\begin{displaymath}
  \frac{\dd}{\dd x}\ln(|W(z)|)
  = \frac{1}{2}\left(\frac{\dd}{\dd z}+\frac{\dd}{\dd\bar z}\right)
  \ln\left(W(z)+\overline{W}(\bar z)\right)
  =\Re\frac{W'(z)}{W(z)}\,,
\end{displaymath}
and analogously,
$$
\frac{\dd}{\dd y}\ln(|W(z)|) =-\Im\frac{W'(z)}{W(z)}\,.
$$

In general, let $\Omega_1$, ..., $\Omega_N$ be mutually disjoint
discrete (possibly empty) sets, and let $\theta_j$,
$j=1,\,\ldots\,,N$, be (not necessarily distinct) real numbers. The
vector potential $\ba$ is defined unambiguously, up to gauge
equivalence, by the expression
\begin{equation}
  \label{s2.1.1}
  \ba=\sum\limits_{j=1}^N\,\theta_j\,\sgrad\ln(|W_j(z)|)
  =\sgrad\ln\!\bigg(\prod\limits_{j=1}^n |W_j(z)|^{\theta_j}\bigg)
\end{equation}
where $W_j$ is an entire function having simple zeros only and with
the zero set being equal to $\Omega_j$. The function $W_j$ differs
from the Weierstrass canonical product related to the set $\Omega_j$
only by a multiplier of the form $\exp(g_j(z))$ where $g_j$ is an
arbitrary entire function. An Aharonov--Bohm potential of the form
(\ref{s2.1.1}) will be called a potential {\em of finite type}. The
operator $H^{\pm}(\ba)$ will be also denoted by the symbols
$H^{\pm}(\Omega_1,\,\ldots,\,\Omega_N;\,\theta_1,\,\ldots,\,\theta_N)$
or $H^{\pm}((\Omega_j)\,;\,(\theta_j))$.

The most important particular cases of potentials of finite type are
those for which the Aharonov--Bohm field is invariant with respect to
a discrete group $\Lambda$ which is formed by motions of the Euclidean
plane $\R^2$ and whose action on $\Omega$ is co-finite. First of all
we shall be interested in the case when the group $\Lambda$ is formed
by parallel translations. Up to isomorphism, there exist just three
groups of this type in the plane and they are characterized by their
rank $r$ ($r=0,\,1,\,2$).

\begin{enumerate}

\item $r=0$. In this case $\Lambda=\{0\}$ and the set $\Omega$ is
  finite.

\item $r=1$. In this case $\Lambda$ is isomorphic to $\Z$ and has the
  form $\Lambda=\{k\omega_0;\,k\in\Z\}$ where $\omega_0$ is a nonzero
  vector from $\R^2$. The set $\Omega$ has the form $\Omega=K+\Lambda$
  where $K$ is a finite subset of the ``elementary strip''
  $F=\{x\in\R^2;\,0\le{}x\cdot\omega_0<|\omega_0|^2\}$ (or, in the
  complex coordinates,
  $F=\{z\in\C;\,0\le\Re\bar{z}\omega_0\,<|\omega_0|^2\}$). Since each
  $\omega\in\Omega$ is uniquely expressible in the form
  $\omega=\kappa+\lambda$, with $\kappa\in{}K$ and
  $\lambda\in\Lambda$, every $\Lambda$-invariant family $\Theta$ is
  unambiguously determined by its subfamily
  $\Theta_{K}=(\theta_\kappa)_{\kappa\in{}K}$.

\item $r=2$. In this case $\Lambda$ is isomorphic to $\Z^2$ and has
  the form $\Lambda=\{k_1\omega_1+k_2\omega_2;\,k_1,k_2\in\Z\}$ where
  $\omega_1$, $\omega_2$ are linearly independent vectors from $\R^2$.
  The set $\Omega$ has the form $\Omega=K+\Lambda$ where $K$ is a
  finite subset of the elementary cell
  $F=\{t_1\omega_1+t_2\omega_2;\,0\le t_1,t_2<1\}$. We shall assume
  that the basis $\omega_1$, $\omega_2$ is positively oriented so that
  $\omega_1\wedge\omega_2=\Im\bar\omega_1\omega_2>0$. This expression
  is nothing but the area $S=S_\Lambda$ of the elementary cell $F$ of
  the lattice $\Lambda$.

\end{enumerate}

\noindent We shall discuss each of these cases separately.

\subsubsection*{Example~5. A finite number of Aharonov--Bohm solenoids}

Let $\Lambda=\{0\}$. In this case the set $\Omega$ is finite,
$\Omega=\{\omega_1,\ldots,\omega_n\}$, and
$$
b=2\pi\sum\limits_{j=1}^n\theta_j\delta(z-\omega_j)\,.
$$
As a rule, the vector potential in this case will be chosen in the
form
$$
\ba=\sum\limits_{j=1}^n\theta_j\,\sgrad\ln(|z-\omega_j|)\,.
$$
The operator $H^{\pm}(\ba)$ will be also denoted by
$H^{\pm}(\omega_1,\,\ldots,\,\omega_n;\,\theta_1,\,\ldots,\,\theta_n)$.

\subsubsection*{Example~6. A chain of Aharonov--Bohm solenoids}

Assume now that the rank of $\Lambda$ equals $1$. Firstly we consider
the case when $K$ contains only one element. Without loss of
generality we assume that $K=\{0\}$. Then $\Omega=\Lambda$,
$\theta_\omega=\theta$ for all $\omega$, and
$$
W_\Omega(z)=
z\prod\limits_{k\in\Z,\,k\ne0}\left(1-\frac{z}{k\omega_0}\right)
e^{z/k\omega_0}=
z\prod\limits_{k=1}^{\infty}\left(1-\frac{z^2}{k^2\omega_0^2}\right)=
\frac{1}{\pi}\sin\!\left(\frac{\pi z}{\omega_0}\right)\,.
$$
Therefore one can set
$$
W(z)=\sin\!\left(\frac{\pi z}{\omega_0}\right)\,.
$$
Consequently,
$$
\ba=\theta\, \sgrad
\ln\!\left(\left|\sin\!\bigg(\frac{\pi z}{\omega_0}\bigg)\right|
\right)\,,
$$
which means that
$$
a_x=\frac{\pi\theta}{\omega_0}\, \Im\ctg\!\left(\frac{\pi
z}{\omega_0}\right)\,,\quad a_y=\frac{\pi\theta}{\omega_0}\,
\Re\ctg\!\left(\frac{\pi z}{\omega_0}\right)\,.
$$

Generally, $\Omega=K+\Lambda$ with an arbitrary finite subset
$K\subset{}F$, and we have
$$
B=\sum\limits_{\kappa\in K}\Phi_\kappa\sum\limits_{\lambda\in\Lambda}
\delta(z-\lambda-\kappa)\,.
$$
Then the vector potential reads
$$
\ba=\sgrad\,\sum\limits_{\kappa\in K}\theta_{\kappa}\,
\ln\!\left(\left|\sin\!\bigg(
    \frac{\pi}{\omega_0}(z-\kappa\bigg)\right|\right)\,.
$$

\subsubsection*{Example~7. A lattice of Aharonov--Bohm solenoids}

Assume now that the rank of $\Lambda$ equals $2$ which means that
$\Lambda$ is a two-dimensional lattice. Again, we shall start from the
case when $K=\{0\}$, hence $\Omega=\Lambda$.  In this case
$W_\Omega(z)$ coincides with the Weierstrass $\sigma$-function of the
lattice $\Lambda$,
$$
\sigma(z;\omega_1,\omega_2)\equiv \sigma(z)
= z\prod\limits_{\omega\in\Omega\setminus\{0\}}
\left(1-\frac{z}{\omega}\right)
\exp\!\left(\frac{z}{\omega}+\frac{z^2}{\omega^2}\right)\,.
$$
At the same time,
$$
\frac{\sigma'(z)}{\sigma(z)}=\zeta(z)=\zeta(z;\omega_1,\omega_2)
$$
is the Weierstrass $\zeta$-function of the lattice $\Lambda$. Thus
$$
\ba=\theta\,\sgrad\,\ln(|\sigma(z)|)=\theta\,\left(\Im\zeta(z),\,
\Re\zeta(z)\right)\,.
$$

In the general case $\Omega=K+\Lambda$ with an arbitrary finite subset
$K\subset F$. Then the magnetic field takes the form
$$
B=\sum\limits_{\kappa\in K}\Phi_\kappa\sum\limits_{\lambda\in\Lambda}
\delta(z-\lambda-\kappa)\,.
$$
One can set
$$
\ba=\sgrad\,\sum\limits_{\kappa\in K}\theta_{\kappa}\,
\ln(\left|\sigma(z-\kappa)\right|)\,.
$$

\begin{rem}
  In all the examples with an Aharonov--Bohm potential of finite type,
  and also in the case of a homogeneous magnetic field, there exists a
  function $\phi(x,y)$ such that $\ba=\sgrad\,\phi$ or, equivalently,
  \begin{equation}
    \label{5.14}
    \Delta\,\phi=b
  \end{equation}
  Namely, one can respectively set in Examples 1, 2, 3, 5, 6, 7:

  $$
  \phi(x,y)=\frac{1}{2}\pi\xi(x^2+y^2)=\frac{1}{2}\pi\xi|z|^2\,,
  $$

  $$
  \phi(x,y)=\frac{1}{2}\theta\,\ln(x^2+y^2)=\theta\,\ln(|z|)\,,
  $$

  $$
  \phi(x,y)=\sum\limits_{j=1}^n \theta_j\,\ln\,|W_j(z)|=
  \ln\!\left(\prod\limits_{j=1}^n|W_j(z)|^{\theta_j}\right),
  $$

  $$
  \phi(x,y)=\sum\limits_{j=1}^n\theta_j\,\ln(|z-\omega_j|)\,,
  $$

  $$
  \phi(x,y)=\sum\limits_{\kappa\in K}
  \theta_\kappa\,\ln\!\left(\left|\sin\!\left(
        \frac{\pi (z-\kappa)}{\omega_0}\right)\right|\right)
  =\ln\!\left(
    \prod\limits_{\kappa\in K}\,\left|\sin\!\left(
        \frac{\pi (z-\kappa)}{\omega_0}\right)
    \right|^{\theta_\kappa}\right),
  $$

  $$
  \phi(x,y)=\sum\limits_{\kappa\in K}
  \theta_\kappa\,\ln(|\sigma(z-\kappa;\omega_1,\omega_2)|)
  =\ln\!\left(\prod\limits_{\kappa\in K}\,
    |\sigma(z-\kappa;\omega_1,\omega_2)|^{\theta_\kappa}\right).
  $$

  Let us note that in the general case when $B$ is a
  $\Lambda$-periodic continuous field the solution of the
  equation~(\ref{5.14}) is expressible in the form
  \begin{equation}
    \label{5.15}
    \phi(z)=\frac{1}{2\pi}\iint\limits_F\ln(|\sigma(z-z')|)\,
    b(z')\,dx'dy'\,,
  \end{equation}
  where $F$ is an elementary cell of the lattice $\Lambda$ \cite{Nov}.
  Actually, we have already seen that
  $$
  \Delta\,\ln(|\sigma(z)|)
  =2\pi\sum\limits_{\lambda\in\Lambda}\delta(z-\lambda)\,.
  $$
  Therefore a formal computation yields
  \begin{equation}
    \label{5.18}
    \Delta\phi(z)=
    \sum\limits_{\lambda\in\Lambda}
    \iint\limits_Fb(z')\,\delta(z-z'-\lambda)\,dx'dy'\,.
  \end{equation}
  For every $z\in\C$ there exists a unique $\lambda_0\in\Lambda$ such
  that $z\in F+\lambda_0$, i.e., $z-\lambda_0\in F$. Then the summands
  in (\ref{5.18}) with $\lambda\ne\lambda_0$ vanish and we have
  $$
  \Delta\phi(z) = \iint\limits_Fb(z')\,
  \delta(z'-(z-\lambda_0))\,dxdy = b(z-\lambda_0) = b(z)\,.
  $$
  In the case of a lattice formed by Aharonov--Bohm solenoids
  formula (\ref{5.15}) makes still sense and it again yields
  $$
  \phi(z)=\theta\,\ln(|\sigma(z)|)\,.
  $$
  Let us note that the Lorentz gauge condition (\ref{1.7b}) follows
  from the equality $\ba=\sgrad\,\phi$.
\end{rem}

In the sequel, the main results will be derived for Hamiltonians
corresponding to three types of systems of Aharonov--Bohm solenoids.
Namely, the set $\Omega$ formed by the intersection points of
solenoids with the plane may be 1) a finite set, 2) a chain or a
finite union of chains, 3) a lattice or a finite union of lattices.
These systems will be called {\em regular}.

\section{A rigorous definition of the Pauli operator as a
  self-adjoint operator}
\label{sec:regdef-orig:3}

Let us return to the symmetric operators $H^{\pm}=H^{\pm}(\ba,\Omega)$
defined in (\ref{1.6a}) while assuming that condition (\ref{1.7}) is
satisfied. Let us introduce the momentum operators
\begin{equation}
  \label{6.5}
  P_x\equiv P_x(\ba,\Omega)=-i\dd_x-a_x\,,\quad
  P_y\equiv P_y(\ba,\Omega)=-i\dd_y-a_y\,.
\end{equation}
In virtue of (\ref{1.7}) these operators can be considered as
symmetric operators in $L^2(\R^2)$ with the domain
$C_0^{\infty}(\R^2\setminus\Omega)$. Following Aharonov--Casher
\cite{AC} we define the operators
\begin{equation}
  \label{6.6}
  T_{\pm}\equiv T_{\pm}(\ba,\Omega)=P_x\pm iP_y\,,
\end{equation}
or $T_+=-2i\dd_{\bar z}-A(z,\bar z)$,
$T_-=-2i\dd_{z}-\bar{A}(z,\bar{z})$ where $A=a_x+ia_y$. Then the
following equalities hold true on $C_0^{\infty}(\R^2\setminus\Omega)$:
\begin{equation}
  \label{6.10}
  T_{+}T_{-}=H^-\,,\quad\quad T_{-}T_{+}=H^+\,.
\end{equation}
By a straightforward computation one can verify a simple but important
lemma.

\begin{lem}\label{orig:lem-3.1}
  The commutation relations
  \begin{equation}
    \label{6.9a}
    [P_x,P_y]=ib\,,\quad [T_-,T_+]=-2b\,,
  \end{equation}
  are valid on $C_0^{\infty}(\R^2\setminus\Omega)$. In particular, if
  $\supp\,B\subset \Omega$ (including the case when $B$ corresponds to
  a system of Aharonov--Bohm solenoids) then the operators $P_x$ and
  $P_y$ (respectively $T_+$ and $T_-$) commute on the domain
  $C_0^{\infty}(\R^2\setminus\Omega)$.
\end{lem}

>From the obvious inclusions
\begin{equation}
  \label{6.12}
  T_{\pm}^*\supset T_{\mp}
\end{equation}
we immediately deduce that {\it the operators $T_{\pm}$ are closable}
and therefore the self-adjoint operators
\begin{equation}
  \label{6.12a}
  H^{\pm}_{\min}\equiv
  H^{\pm}_{\min}(\ba,\Omega)=T_{\pm}^*\overline T_{\pm}
\end{equation}
are well defined (see, e.g., \cite[Theorem X.25]{RSII}).
The associated quadratic forms $h^{\pm}_{\min}$ are closures of
positive forms defined on $C_0^{\infty}(\R^2\setminus\Omega)$ by the
expressions
\begin{equation}
  \label{6.12b}
  \bra T_{\pm}\phi|T_{\pm}\psi\ket\,,
\end{equation}
respectively.

On the other side, let us consider a quadratic form defined on
$C_0^{\infty}(\R^2\setminus\Omega)$ by the relation
\begin{equation}
  \label{6.12c}
  s^{\pm}(\phi,\psi)=\bra P_x\phi|P_x\psi\ket+\bra P_y\phi|P_y\psi\ket
  \mp\bra b\phi|\psi\ket\,.
\end{equation}

By a straightforward computation using relation (\ref{6.9a}) one can
show the following lemma.

\begin{lem}\label{orig:lem-3.2}
  The quadratic forms $h_{\min}^{\pm}$ and $s^{\pm}$ coincide on
  $C_0^{\infty}(\R^2\setminus\Omega)$.
\end{lem}

In particular, if the support of $B$ is contained in $\Omega$ then the
quadratic forms $h_{\min}^+$ and $h_{\min}^-$ coincide on
$C_0^{\infty}(\R^2\setminus\Omega)$ and therefore they are necessarily
equal.

\begin{cor}
  If $B$ is a distribution with a support contained in $\Omega$ then
  the operators $H_{\min}^+$ and $H_{\min}^-$ coincide. In particular,
  for the vector potential $\ba$ of a system of the Aharonov--Bohm
  vortices supported on a set $\Omega\subset\R^2$ these operators
  coincide with the Friedrichs extension of the symmetric operator
  defined on $C^\infty_0(\R^2\setminus\Omega)$ by the differential
  expression
\begin{equation}
                         \label{cor.lem.4}
H_0(\ba)=(i\dd_x+a_x)^2+(i\dd_y+a_y)^2\,.
\end{equation}
\end{cor}

In view of Lemma~\ref{orig:lem-3.2} we shall sometimes simply
write $H_{\min}$ instead of $H_{\min}^{\pm}$. The operator
$H_{\min}$ has been investigated in detail in \cite{MOR}.

 Jointly with the operator $H_{\min}(\ba,\Omega)$ let us
consider the operators
\begin{equation}
  \label{6.12d}
  H^{\pm}_{\max}\equiv H^{\pm}_{\max}(\ba,\Omega)=
  \overline T_{\mp}T_{\mp}^*
\end{equation}
with the associated quadratic forms defined on
${\cal{D}}(H_{\max}^{\pm})$ by the expressions
\begin{equation}
  \label{6.12e}
  h^{\pm}_{\max}(\phi,\psi)=\bra T_{\mp}^*\phi|T_{\mp}^*\psi\ket\,,
\end{equation}
respectively.

The definitions of $H^{\pm}_{\max}(\ba,\Omega)$ and
$H^{\pm}_{\min}(\ba,\Omega)$ in principle depend on the choice of the
discrete set $\Omega$. If $\Omega$ coincides with the singular support
of $b$, however, we shall simply write, similarly as in
Section~\ref{sec:pauliop-orig:1}, $H^{\pm}_{\min}(\ba)$ and
$H^{\pm}_{\max}(\ba)$ since in that case the vector potential $\ba$
determines $\Omega$ unambiguously.

If the field $B$ is sufficiently regular and $\Omega=\emptyset$ then
the operator $H^{\pm}_{\min}$ coincides with the operator
$H^{\pm}_{\max}$ \cite{CFKS}. This is not true, however, for operators
with Aharonov--Bohm fluxes (see \cite{Ara1}, \cite{MOR}). Since in
this case $H^{\pm}_{\min}$ is defined by expression (\ref{cor.lem.4})
and is independent of spin, this operator is the Schr\"odinger
operator of a spinless particle in the presence of the Aharonov--Bohm
fluxes (or the Schr\"odinger operator of a particle with spin when
interaction of the spin with the field can be neglected). On the other
hand, $H^{\pm}_{\min}$ are defined by expression (\ref{1.6a}), they
depend on the spin and may be considered as the Pauli operators for an
electron with the gyromagnetic ratio $g=2$. Below we are interesting
in the properties of ground states of the operator $H^{\pm}_{\max}$.

For the analysis of operators $H^{\pm}_{\max}$ the following
description of the operators $T^*_{\pm}$ will be useful. Namely, owing
to condition (\ref{1.7}) the differential operators $-i\partial_x-a_x$
and $-i\partial_y-a_y$ are well defined on the space of distributions
${\cal{D}}'(\R^2\setminus\Omega)$. Consequently, the operators
$T_{\pm}$ defined on $C_0^{\infty}(\R^2\setminus\Omega)$ can be
naturally extended to linear mappings $\widetilde{T}_{\pm}$ defined on
${\cal{D}}'(\R^2\setminus\Omega)$. Using the fact that $L^2(\R^2)$ is
naturally embedded into ${\cal{D}}'(\R^2\setminus\Omega)$ we get the
following lemma.

\begin{lem}
  The operator $T^*_{\pm}$ is a restriction of $\widetilde{T}_{\mp}$
  to the domain
  $$
  \{f\in L^2(\R^2);\,\widetilde T_{\mp}f\in L^2(\R^2)\}\,.
  $$
\end{lem}

Using this observation we can prove the following lemma.

\begin{lem}\label{orig:lem-3.4}
  Let $C$ be the operator of complex conjugation, $Cf=\bar f$. Then
  $CH_{\max}^{\pm}(\ba,\Omega)=H_{\max}^{\mp}(-\ba,\Omega)C$ and
  $CH_{\min}^{\pm}(\ba,\Omega)=H_{\min}^{\mp}(-\ba,\Omega)C$.
\end{lem}

\begin{cor}\label{orig:coroflem-3.4}
  The operators $H_{\max}^{+}(\ba,\Omega)$ and
  $H_{\max}^{-}(-\ba,\Omega)$ have the same spectra. In particular,
  they have the same eigenvalues with equal multiplicities. An
  analogous proposition holds true for the couple of operators
  $H_{\min}^{+}(\ba,\Omega)$ and $H_{\min}^{-}(-\ba,\Omega)$.
\end{cor}

\section{Elimination of Aharonov--Bohm solenoids with integer
  fluxes}
\label{sec:intflux-orig:4}

In this section we consider a vector potential $\tilde\ba$ of the form
$$
\tilde\ba=\ba+\ba_{AB}
$$
where $\ba_{AB}$ is a vector potential corresponding to a system
of Aharonov--Bohm solenoids intersecting the plane in the points
of $\Omega$. We describe here briefly the "gauge-periodicity" of
the operators with the vector potential $\tilde\ba$; details can
be found e.g. in \cite{Hel,HHN,HHHO,HHHO1}.

First we shall assume that the considered solenoids carry {\it equal}
fluxes of the value $\theta_{AB}$. In this case we set
$\ba_{AB}=\theta_{AB}\,\sgrad\,\ln(|W(z)|)$ (cf. Example~4 in
Section~\ref{sec:examples-orig:2}). Let $\theta_{AB}$ be an integer.
Then the function
$$
g(z,\bar z)=\exp\!\left(\theta_{AB}\,
\ln\!\left(\frac{W(z)}{|W(z)|}\right)\right)
=\exp\!\big(i\theta_{AB}\,\arg(W(z))\big)\,,
$$
is well defined and continuous in the domain $\C\setminus\Omega$.
Clearly, $|g(z,\bar z)|=1$, $\forall{}z\in\C\setminus\Omega$, and,
moreover, $g\in{}C^{\infty}(\R^2\setminus\Omega)$.

\begin{lem}\label{orig:lem-4.1}
  If $\theta_{AB}$ is an integer then the following relations hold
  true
  $$
  g^{-1}P_x(\tilde\ba,\Omega)g=P_x(\ba,\Omega)\,,\quad
  g^{-1}P_y(\tilde\ba,\Omega)g=P_y(\ba,\Omega)\,.
  $$
\end{lem}

\begin{proof}
  It suffices to show that
  \begin{equation}
    \label{0.1}
    -i\,\grad g=g\,\ba_{AB}\,.
  \end{equation}
  Actually, we have
  \begin{eqnarray*}
    \frac{\dd}{\dd x}\,\ln\!\left(\frac{W(z)}{|W(z)|}\right)
    &=& \left(\frac{\dd}{\dd z}+\frac{\dd}{\dd\bar z}\right)
    \left(\ln(W(z))-\frac{1}{2}
      \left(\ln(W(z))+\ln(\overline{W(z)})\right)\right) \\
    &=& i\,\Im\frac{W'(z)}{W(z)} = i\,\theta_{AB}^{-1}\,
    a_{AB,x}\,.
  \end{eqnarray*}
  Analogously,
  \begin{displaymath}
    \frac{\dd}{\dd y}\ln\!\left(\frac{W(z)}{|W(z)|}\right)
    = i\,\Re\frac{W'(z)}{W(z)}=i\,\theta_{AB}^{-1}\,a_{AB,y}\,.
  \end{displaymath}
  Relation (\ref{0.1}) obviously follows from these equalities.
\end{proof}

Assume now that $\ba_{AB}$ is a vector potential corresponding to an
Aharonov--Bohm field of finite type whose singular support coincides
with $\Omega=\Omega_1\cup\ldots\cup\Omega_n$, and with an array of
fluxes denoted by $\Theta=(\theta_j)_{1\le j\le n}$. Let
$(m_j)_{1\le{}j\le n}$ be an arbitrary array of integers and let
$\tilde\ba_{AB}$ be another Aharonov--Bohm potential of finite type
defined by the same array of sets $(\Omega_j)$ but with the fluxes
$\tilde\theta_j=\theta_j+m_j$.

\begin{thm}\label{orig:thm-4.1}
  Assume that $\ba$, $\ba_{AB}$ and $\tilde\ba_{AB}$ have the same
  meaning as described above. Then
  $H^{\pm}_{\min}(\ba+\ba_{AB},\Omega)$ (respectively,
  $H^{\pm}_{\max}(\ba+\ba_{AB},\Omega)$) is unitarily equivalent to
  the operator $H^{\pm}_{\min}(\ba+\tilde\ba_{AB},\Omega)$
  (respectively, $H^{\pm}_{\max}(\ba+\tilde\ba_{AB},\Omega)$).
\end{thm}

\begin{proof}
  Let $T_{1\pm}$, $T_{2\pm}$ be the operators corresponding to the
  vector potentials $\ba+\ba_{AB}$ and $\ba+\tilde\ba_{AB}$,
  respectively, as described in Section~\ref{sec:regdef-orig:3}. By
  construction,
  \begin{displaymath}
    {\cal D}(T_{1\pm})={\cal D}(T_{2\pm})
    = C_0^{\infty}(\R^2\setminus\Omega)\,.
  \end{displaymath}
  Applying repeatedly Lemma~\ref{orig:lem-4.1} one can show that there
  exists a unitary operator $U$ such that
  \begin{displaymath}
    U\!\left(C_0^{\infty}(\R^2\setminus\Omega)\right)
    =C_0^{\infty}(\R^2\setminus\Omega)
  \end{displaymath}
  and
  \begin{displaymath}
    U^{-1}T_{2\pm}U=T_{1\pm}.
  \end{displaymath}
  From the unitarity of $U$ it follows that
  \begin{displaymath}
    U^{-1}\overline{T}_{2\pm}U=\overline{T}_{1\pm}
    \text{ and }
    U^{-1}T_{2\pm}^\ast U=T_{1\pm}^\ast.
  \end{displaymath}
  Consequently,
  \begin{displaymath}
    U^{-1}H^{\pm}_{\min}(\ba+\tilde\ba_{AB},\Omega)U
    = U^{-1}T_{2\pm}^\ast\overline{T}_{2\pm}U
    = T_{1\pm}^\ast\overline{T}_{1\pm}
    = H^{\pm}_{\min}(\ba+\ba_{AB},\Omega)
  \end{displaymath}
  and
  \begin{displaymath}
    U^{-1}H^{\pm}_{\max}(\ba+\tilde\ba_{AB},\Omega)U
    = U^{-1}\overline{T}_{2\mp}T_{2\mp}^\ast U
    = \overline{T}_{1\mp}T_{1\mp}^\ast
    = H^{\pm}_{\max}(\ba+\ba_{AB},\Omega).
  \end{displaymath}
  This shows the theorem.
\end{proof}

\begin{cor}\label{orig:cor-1}
  If all fluxes $\theta_j$ are integers then the operator
  $H^{\pm}_{\min}(\ba+\ba_{AB},\Omega)$ (respectively,
  $H^{\pm}_{\max}(\ba+\ba_{AB},\Omega)$) is unitarily equivalent to
  the operator $H^{\pm}_{\min}(\ba,\Omega)$ (respectively,
  $H^{\pm}_{\max}(\ba,\Omega)$).
\end{cor}

Let us formulate separately two most important cases of this
corollary. The first one is based on the fact that the both operators
$H^{\pm}_{\min}(0,\Omega)$ and $H^{\pm}_{\max}(0,\Omega)$ do not
depend on the choice of the discrete set $\Omega$ and coincide with
the Laplace operator $-\Delta$.

\begin{cor}\label{orig:cor-2}
  Let $\Omega$ be a discrete set which is invariant with respect to a
  co-finite action of a lattice $\Lambda$ of rank $r$, $0\le r\le 2$.
  Assume that $\ba=0$ and $\ba_{AB}$ is a vector potential
  corresponding to a system of Aharonov--Bohm solenoids supported on
  the set $\Omega$ and such that all fluxes are integers. Then each of
  the operators $H^{\pm}_{\min}(\ba_{AB},\Omega)$ and
  $H^{\pm}_{\max}(\ba_{AB},\Omega)$ is unitarily equivalent to the
  Laplace operator $-\Delta$.
\end{cor}

\begin{proof}
  Since the action is co-finite we are again in the situation when
  $\Omega$ splits into a finite union
  $\Omega=\Omega_1\cup\ldots\cup\Omega_n$. Hence one can apply
  Corollary~\ref{orig:cor-1}. The unitary operator induced by
  multiplication with the function $g$ acts locally in the form sense
  \cite{Sho} and therefore each of the operators
  $H^{\pm}_{\min}(0,\Omega)$ and $H^{\pm}_{\max}(0,\Omega)$ is a point
  perturbation of $-\Delta$ supported on the set $\Omega$. The
  perturbed operator is clearly positive and local in the form sense
  \cite{Pan}. On the other hand, every nontrivial point perturbation
  in the two-dimensional case is known to have a strictly negative
  infimum of the quadratic form over unit vectors \cite{AGHH}.
\end{proof}

Since the minimum of spectrum in the case of a periodic point
perturbation of the Landau operator is strictly smaller than the
minimum of spectrum of the unperturbed operator \cite{Gey} the
following corollary is also true.

\begin{cor}\label{orig:cor-3}
  Let $\ba$ be a vector potential of a nonzero homogeneous magnetic
  field and assume again that the discrete set $\Omega$ is invariant
  with respect to a co-finite action of a lattice $\Lambda$ of rank
  $r$, $0\le r\le 2$. Then for $b>0$, each of the operators
  $H^{+}_{\min}(\ba+\ba_{AB},\Omega)$ and
  $H^{+}_{\max}(\ba+\ba_{AB},\Omega)$ is unitarily equivalent to the
  Landau operator $H^{+}(\ba)$. For $b<0$, an analogous statement is
  true for the operators $H^{-}_{\min}(\ba+\ba_{AB},\Omega)$ and
  $H^{-}_{\max}(\ba+\ba_{AB},\Omega)$.
\end{cor}

To simplify the discussion to follow we shall assume once for all that
an appropriate gauge transformation has been applied so that the
values of all involved Aharonov--Bohm fluxes belong to the interval
$[0,1[\,$. If there are some zero values then $\Omega$ is strictly
larger then the singular support of $b$. As shown by
Corollaries~\ref{orig:cor-2} and \ref{orig:cor-3}, the zero values can
be eliminated in some particular cases. We shall proceed in our
simplifications even further. If not said otherwise, we assume
everywhere in what follows that {\em the values of Aharonov--Bohm
  fluxes belong to the interval $]0,1[$ and, consequently, the
  singular support of $b$ coincides with $\Omega$}.

\section{The ground states (zero modes) of the Pauli operator}
\label{sec:groundsts-orig:5}

It follows immediately from the definition of the operators
$H^{\pm}_{\max}$ and $H^{\pm}_{\min}$ that they are nonnegative.
Consequently, if the equation
\begin{equation}
  \label{7.3}
  H^{\pm}_{\min}\psi=0
\end{equation}
or the equation
\begin{equation}
  \label{7.3a}
  H^{\pm}_{\max}\psi=0\,
\end{equation}
has a solution in $L^2(\R^2)$ then this solution $\psi_{\pm}$ (called
{\em zero mode}) is a ground state of the corresponding operator.
Since the equality $H^{\pm}_{\min}\psi=0$ implies
$\bra{}H^{\pm}_{\min}\psi|\psi\ket=0$, i.e., the equality
$\|\overline{T}_{\pm}\psi\|^2=0$, equation (\ref{7.3}) is equivalent
to the equality
\begin{equation}
  \label{7.3b}
  \overline{T}_{\pm}\psi=0\,.
\end{equation}
Analogously, equation (\ref{7.3a}) is equivalent to the equality
\begin{equation}
  \label{7.3c}
  T^*_{\mp}\psi=0\,,
\end{equation}
or, this is the same, to the condition
\begin{equation}
  \label{7.3d}
  \widetilde T_{\pm}\psi=0\,,\quad \psi\in L^2(\R^2)\,.
\end{equation}

Suppose that the vector potential $\ba$ was chosen to have the form
$\ba=\sgrad\,\phi$ where $\phi$ satisfies the equation $\Delta\phi=b$
in the sense of distributions. We shall seek a solution of equation
(\ref{7.3d}) in the form
\begin{equation}
  \label{7.4}
  \psi_{\pm}(x,y)=\exp(\mp\phi(x,y))f(x,y)
  =\exp(\mp\phi(z,\bar{z}))f(z,\bar z)\,,
\end{equation}
where $f$ has to be chosen so that $\psi_{\pm}\in L^2(\R^2)$ ({\it the
  Aharonov--Casher ansatz}). In the space of distributions
${\cal{D}}'(\R^2\setminus\Omega)$ we have
\begin{eqnarray*}
  T^*_-\psi_+ &=& \widetilde T_+\psi_+ \nonumber\\
  &=& \exp(-\phi)\left(\big(i(\dd_x\phi-a_y)f
    +(-\dd_y\phi-a_x)f\big)-
    i(\dd_xf+i\dd_yf)\right) \nonumber\\
  &=& -2i\exp(-\phi)\frac{\dd f}{\dd\bar z}
\end{eqnarray*}
and
\begin{eqnarray*}
  T^*_+\psi_- &=& \widetilde T_-\psi_- \nonumber\\
  &=& \exp(\phi)\left(\big(i(-\dd_x\phi+a_y)f
    +(-\dd_y\phi-a_x)f\big)
    -i(\dd_xf-i\dd_yf)\right) \nonumber\\
  &=& -2i\exp(\phi)\frac{\dd f}{\dd z}\,.
\end{eqnarray*}
>From here we deduce that the relation
\begin{equation}
  \label{7.7}
  H^+_{\max}\psi_+=0\,,\quad \psi_+\in L^2(\R^2),
\end{equation}
is equivalent to the condition
\begin{equation}
  \label{7.7a}
  \frac{\dd f}{\dd\bar z}=0\,\,\,(z\in\C\setminus\Omega)\,,\quad
  \exp(-\phi)f\in L^2(\R^2)\,.
\end{equation}
Analogously, the relation
\begin{equation}
  \label{7.8}
  H^-_{\max}\psi_-=0\,,\quad \psi_-\in L^2(\R^2),
\end{equation}
is equivalent to the condition
\begin{equation}
  \label{7.8a}
  \frac{\dd f}{\dd z}=0\,\,\,(z\in\C\setminus\Omega)\,,\quad
  \exp(\phi)f\in L^2(\R^2)\,.
\end{equation}
This shows the following theorem due to Aharonov and Casher \cite{AC}.

\begin{thm}\label{orig:thm-5.1}
  Assume that a vector potential $\ba$ is expressed in the form
  $\ba=\sgrad\,\phi$ where $\phi$ satisfies the equation
  $\Delta\phi=b$ in the sense of distributions. Then solutions of the
  equation $H^+_{\max}\psi=0$ in $L^2(\R^2)$ are exactly those
  functions from $L^2(\R^2)$ which have the form $\psi_+(z,\bar
  z)=\exp(-\phi(z,\bar z))f(z)$ where $f$ is a holomorphic function in
  the domain $\C\setminus\Omega$.

  Similarly, solutions of the equation $H^-_{\max}\psi=0$ in
  $L^2(\R^2)$ are exactly those functions from $L^2(\R^2)$ which have
  the form $\psi_-(z,\bar{z})=\exp(\phi(z,\bar{z}))f(\bar z)$ where
  $f$ is a holomorphic function in the domain $\C\setminus\Omega$.
\end{thm}

Let us point out an interesting consequence of the theorem.

\begin{prop}
  Assume that the both operators $H_{\max}^+$ and $H_{\max}^-$ have
  zero modes. Then they are distinct. In particular, the set
  $C^{\infty}_0(\C\setminus\Omega)$ is not a core for at least one of
  them.
\end{prop}

\begin{proof}
  Let $\psi$ be a zero mode of $H_{\max}^+$. Suppose that this
  operator coincides with $H_{\max}^-$. Then $\psi$ is a zero mode for
  $H_{\max}^-$ as well. Using notation of Theorem~\ref{orig:thm-5.1}
  we have $\psi=\exp(-\phi)f=\exp(\phi)g$ where $f$ is holomorphic in
  the domain $\C\setminus\Omega$ and $g$ is antiholomorphic in the
  same domain. Since $\phi$ is real it holds true that
  $|\psi|^2=f\bar{g}$. Taking into account that $\bar g$ is
  holomorphic the last equality implies that $f\bar g$ is a constant
  function and hence the same is true for $|\psi|^2$.  Since
  $\psi\in{}L^2(\R^2)$ it follows that $\psi=0$, a contradiction.
\end{proof}

\section{Zero modes of the operators $H_{\max}^{\pm}$ with
  Aharonov--Bohm potential of finite type}
\label{sec:zeromod-orig:6}

\subsection{Formulation of the problem}
\label{subsec:form-prob}

In this section we shall study ground states of the operator
$H_{\max}^{\pm}(\ba,\Omega)$ for an Aharonov--Bohm potential $\ba$ of
finite type determined by mutually disjoint discrete sets
$\Omega_1\,,\ldots, \Omega_n$ such that
$\Omega=\Omega_1\cup\ldots\cup\Omega_n$, and by fluxes (not
necessarily distinct) $\theta_1\,,\ldots, \theta_n$ (cf. Example~4
from Section~\ref{sec:examples-orig:2}). Recall that we assume that
$0<\theta_j<1$, for all $j$.

We can rephrase the formulation of the problem. Namely, according to
Theorems~\ref{orig:thm-4.1} and \ref{orig:thm-5.1} we have to study
square integrability of a function $\psi$ having the form
\begin{equation}
  \label{6.1.0}
  \psi(z,\bar z)=f(z)\prod\limits_{j=1}^n\,|W_j(z)|^{-\theta_j+m_j}
\end{equation}
where the numbers $m_j$ are integers, $f(z)$ is holomorphic or
antiholomorphic in the domain $\C\setminus\Omega$, and the functions
$W_j$ determine the potential $\ba$ according to formula
(\ref{s2.1.1}).

In this section the following lemma will be useful.

\begin{lem}\label{orig:lem-6.1}
  Assume that a function $f(z)$ is expressible in an annulus
  $r_1<|z|<r_2$ as a Laurent series
  $$
  f(z)=\sum\limits_{n=-\infty}^{\infty}\,a_nz^n\,.
  $$
  Then for $r_1<r<r_2$ and arbitrary $n\in\Z$ it holds true that
  \begin{equation}
    \label{6.1.1}
    \int\limits_{|z|=r}|f(z)|\,|dz| \ge 2\pi |a_n|r^{n+1}\,.
  \end{equation}
  In particular, if $r_1=0$ and $n>-2$ then
  \begin{equation}
    \label{6.1.2}
    \iint\limits_{|z|<r}|f(z)|\,dxdy \ge
    \frac{2\pi |a_n|}{n+2}r^{n+2}\,,
  \end{equation}
  if $r_1=0$ and $a_n\ne 0$ for some $n\le-2$ then
  \begin{equation}
    \label{6.1.3}
    \iint\limits_{|z|<r}|f(z)|\,dxdy = \infty\,,
  \end{equation}
  if $r_2=\infty$ and $a_n\ne 0$ for some $n\ge-2$ then
  \begin{equation}
    \label{6.1.4}
    \iint\limits_{|z|>r}|f(z)|\,dxdy = \infty\,.
  \end{equation}
\end{lem}

\begin{proof}
  The proof immediately follows from the simple estimate
  $$
  \int\limits_{|z|=r}|f(z)|\,|dz|
  =r\int\limits_0^{2\pi}|f(re^{i\phi})|\,d\phi \ge
  r\left|\int\limits_0^{2\pi}e^{-in\phi}f(re^{i\phi})\,d\phi\right|
  =2\pi |a_n|r^{n+1}\,.\textrm{ }\qed
  $$
  \renewcommand{\qed}{}
\end{proof}

An application of Lemma~\ref{orig:lem-6.1} yields the following
auxiliary result.

\begin{lem}\label{orig:lem-6.2}
  Assume that in (\ref{6.1.0}) it holds $m_j=0$, for all $j$, i.e.,
  $$
  \psi(z,\bar z)=f(z)\prod\limits_{j=1}^n\,|W_j(z)|^{\beta_j}
  $$
  where $-1<\beta_j<0$, for each $j$, and $f(z)$ is holomorphic in
  the domain $\C\setminus\Omega$. If $\psi\in L^2(\R^2)$ then $f$ is
  an entire function.
\end{lem}

\begin{proof}
  Assume that $\omega\in\Omega_k$ and $r>0$ is sufficiently small so
  that $D(\omega,r)\cap\Omega=\{\omega\}$ where $D(\omega,r)$ is the
  disc with radius $r$ centered at the point $\omega$. Since
  $W_j(\omega)\ne 0$, for $j\ne{}k$, and $\omega$ is a simple zero of
  $W_k(z)$ one can assume that $r$ is small enough so that it holds
  $$
  \prod\limits_{j=1}^n\,|W_j(z)|^{\beta_j}\ge c > 0,
  $$
  for $0<|z-\omega|<r$.

  Now one can show that $\omega$ cannot be a pole nor an essential
  singularity of $f$. Otherwise $\omega$ would be a pole of even order
  or an essential singularity of $f^2$. In any case
  Lemma~\ref{orig:lem-6.1} implies that
  $$
  \iint\limits_{D(\omega,r)}|\psi(z,\bar z)|^2\,dxdy \ge
  c^2\iint\limits_{D(\omega,r)}|f(z)|^2\,dxdy =\infty\,,
  $$
  a contradiction.
\end{proof}

\subsection{Finite number of Aharonov--Bohm fluxes}

We start from the simplest case, i.e., from the Hamiltonian
$H_{\max}^{\pm}(\Omega,\Theta)$ corresponding to a finite number of
Aharonov--Bohm fluxes (Example~5 from
Section~\ref{sec:examples-orig:2}). Then
$\Omega=\{\omega_1,\ldots,\omega_n\}$ is a finite set,
$\Theta=(\theta_1,\ldots,\theta_n)$, $0<\theta_j<1$. In this case zero
modes may occur under suitable assumptions on fluxes $\theta_j$. More
precisely, the following theorem is true \cite{Ara1}.

\begin{thm}\label{orig:thm-6.1}
  A sufficient and necessary condition for the operators
  $H_{\max}^+(\Omega,\Theta)$ and $H_{\max}^-(\Omega,\Theta)$ to have
  zero modes is
  \begin{equation}
    \label{eq:sum_tj_gt_1}
    \sum\limits_{j=1}^n\theta_j > 1\,,
  \end{equation}
  in the former case, and
  \begin{equation}
    \label{eq:sum_tj_lt_J-1}
    \sum\limits_{j=1}^n\theta_j < n-1
  \end{equation}
  in the latter case.
\end{thm}

\begin{proof}
  Let us start from the operator $H_{\max}^+$. We have to find a
  nonzero function $f$ which is holomorphic in the domain
  $\C\setminus\Omega$ and such that the function
  \begin{equation}
    \label{61.1}
    \psi(z,\bar z)=f(z)\prod\limits_{j=1}^n|z-\omega_j|^{-\theta_j}
  \end{equation}
  is square integrable. Suppose that condition~(\ref{eq:sum_tj_gt_1})
  is satisfied. Taking for $f$ a constant function it is easy to
  verify that in that case we get a square integrable function $\psi$.

  Conversely, assume that $\psi$ is square integrable but
  condition~(\ref{eq:sum_tj_gt_1}) is false. Then from (\ref{61.1})
  one easily deduces that $f$ cannot be a nonzero constant.
  Furthermore, from the equality
  $$
  f(z)=\psi(z,\bar z)\,\prod\limits_{\theta_j\in J}
  |z-\omega_j|^{\theta_j}\,,
  $$
  we find that there exists a constant $c_1>0$ such that
  $$
  |f(z)|\le c_1 (1+|z|)|\psi(z,\bar z)|\,\textrm{ on }\C.
  $$
  Consequently, if $r>1$ then
  $$
  \iint\limits_{|z|<r}|f(z)|^2\,dxdy \le c_2r^2
  $$
  where
  $$
  c_2 = 4\,c_1^2\iint\limits_{\R^2}|\psi(z,\bar{z})|^2\,dxdy\,.
  $$
  From inequality~(\ref{6.1.1}) in Lemma~\ref{orig:lem-6.1} it
  follows that $f(z)$ is a constant function. This contradiction
  proves the theorem in the case of the operator $H_{\max}^+$. To
  prove the theorem in the case of the operator $H_{\max}^-$ one can
  either modify the above argument or to apply
  Corollary~\ref{orig:coroflem-3.4}.
\end{proof}

\subsection{A chain of Aharonov--Bohm fluxes}

Here we show that the Hamiltonians $H_{\max}^{\pm}$ corresponding to a
finite union of chains of Aharonov--Bohm fluxes have infinitely many
zero modes. In this subsection we use the notation from Example~6 in
Section~\ref{sec:examples-orig:2}.

The proof uses the following elementary estimate. Since for $z=x+iy$,
$x,\,y\in\R$, we have
$|\sin(z)|^2=\ch^2(y)-\cos^2(x)=\sh^2(y)+\sin^2(x)$, it holds true
that $|\sh(y)|\le|\sin(z)|\le\ch(y)$. Hence
\begin{equation}
  \label{6.4a}
  \frac{e^{|y|}-1}{2}\le|\sin(z)|\le e^{|y|}\,.
\end{equation}

\begin{thm}
  \label{orig:thm-6.3}
  Let a uniformly discrete set $\Omega$ be expressible as a disjoint
  union of a finite number of chains $\Omega_1\,,\ldots,\Omega_n$, and
  let the chain $\Omega_j=K_j+\Lambda_j$ carry Aharonov--Bohm fluxes
  $(\theta_\kappa)_{\kappa\in{}K_j}$ ($j=1\,,\ldots,\,n$,
  $0<\theta_\kappa<1$). Then the Hamiltonians $H_{\max}^{\pm}(\Omega)$
  have infinitely degenerate zero modes.
\end{thm}

\begin{proof}
  In the proof we shall consider only the operator $H_{\max}^+$.
  Using Lemma~\ref{orig:lem-9.1} we may assume that each chain
  $\Omega_j$ is contained in a line $L_j$ and that it holds
  $L_j\ne{}L_k$ for $j\ne k$. Then the Bravais lattice $\Lambda_j$ of
  the chain $\Omega_j$ has the form $\Lambda_j=\{k\omega_j;\textrm{
  }k\in\Z\}$, $\omega_j\in\C$, $\omega_j\ne 0$. Without loss of
  generality we can suppose that $\omega_1>0$ and $\kappa_1=0$. Hence
  $L_1=\R$ and $L_j=\omega_j\R+\kappa_j$ ($j=2,\ldots,n$) where
  $\kappa_j$ is a fixed element from $K_j$.

  For each line $L_j$ we shall construct a strip $P_j$ with border
  lines parallel to $L_j$ and containing $L_j$ in its interior.
  Furthermore, let $Q$ be a sufficiently large disk centered at 0 such
  that outside $Q$ the strips $P_j$ do not intersect each other.

  It suffices to show that there exists an infinite number of linearly
  independent entire functions $f(z)$ for which the function
  $$
  \psi(z,\bar z)= f(z)g_1(z,\bar z)\cdot\ldots\cdot g_n(z,\bar z),
  $$
  with
  \begin{equation}
    \label{6.3.1}
    g_j(z, \bar z)= \prod_{\kappa\in K_j}
    \left|\sin\!\left(\frac{\pi(z-\kappa)}
        {\omega_j}\right)\right|^{-\theta_\kappa},
  \end{equation}
  is square integrable. We shall show that this condition is satisfied
  for any function
  $$
  f(z) = \frac{\sin(\alpha z)}{z}
  $$
  where
  \begin{equation}
    \label{eq:cond-u-chains}
    0<\alpha< \theta := \frac{\pi}{\omega_1}
    \sum\limits_{\kappa\in  K_1}\theta_\kappa\,.
  \end{equation}

  The verification follows from a series of claims.

  \medskip

  \noindent{\rm(A)} $\psi\in L^2(Q)$,

  \medskip

  \noindent{\rm(B)} {\it each function $g_j$ is bounded outside
    the strip $P_j$},

  \medskip

  \noindent{\rm(C)} {\it the function $g_1(z,\bar z)\sin(\alpha z)$
    is bounded outside the strip $P_1$}.

  \medskip

  Claim {\rm(A)} follows from the fact that $f$ is bounded on $Q$ and
  that the functions $g_j$ have square integrable singularities.
  Moreover, only finitely many singularities are contained in $Q$.
  Claims {\rm(B)} and {\rm(C)} are consequences of the inequalities in
  (\ref{6.4a}) and condition~(\ref{eq:cond-u-chains}).

  To complete the proof it remains to show that

  \medskip

  \noindent{\rm(D)} $\psi\in L^2(P_j\setminus Q)$,
  $\forall\,j=1,\ldots,n$,

  \medskip

  \noindent{\rm(E)} $\psi\in
  L^2(\R^2\setminus(P_1\cup\ldots\cup P_n))$.

  \medskip

  To show Claim~{\rm(D)} notice that {\rm(B)} and {\rm(C)} imply the
  estimate
  $$
  |\psi(z,\bar z)|\le \frac{c_j}{|z|}\,|g_j(z,\bar z)|,
  $$
  valid on $P_j\setminus Q$ with some constant $c_j>0$, and that
  the function $g_j(z,\bar z)$ is periodic along the line $L_j$. For
  $j=1$ one uses also that $\sin(\alpha{}z)$ is bounded on the strip
  $P_1$.

  To show Claim~{\rm(E)} let us point out that the inequality
  $$
  |\psi(z,\bar z)|\le c'\left|
    \frac{\sin(\alpha z)}{z}\right||g_1(z,\bar z)|
  $$
  holds true on $\C\setminus(P_2\cup\ldots\cup P_n)$ with some
  constant $c'>0$, as it follows from {\rm(B)}. From inequalities
  (\ref{6.4a}) one derives that
  $$
  |\psi(z,\bar z)|\le c''\,
  \frac{\exp\big((\alpha-\theta)|y|\big)}{|z|}\,.
  $$
  on $\C\setminus(P_1\cup{}P_2\cup\ldots\cup P_n)$. Finally,
  condition~(\ref{eq:cond-u-chains}) implies that
  $\psi\in{}L^2(\R^2\setminus(P_1\cup\ldots\cup{}P_n))$.
\end{proof}

Under more restrictive conditions on the fluxes $\theta_\kappa$ the
assumption on the uniform discreteness can be dropped. Since every
chain is a union of one-atom chains we can confine ourselves to such
chains. Moreover, it is clear that a union of chains need not be a
uniformly discrete set only in the case when among the chains in
question there are at least two contained in the same line.
Consequently, it suffices to analyze the case when the chains are
contained in a single line, say, in the real axis $\R$. Suppose that
$\Omega_j=\kappa_j+\Lambda_j$ where $\Lambda_j=\{\omega_j{}k;\textrm{
}k\in\Z\}$, $\kappa_j\in\R$, $\omega_j> 0$ ($j=1,\ldots,n$), are
mutually disjoint one-atom chains and $\theta_1,\ldots,\theta_n$
($0<\theta_j<1$) are the corresponding Aharonov--Bohm fluxes.

\begin{thm}\label{orig:thm-6.4}
  Assume that all chains $\Omega_j$, $j=1,\ldots,n$, are contained in
  $\R$. Then the Hamiltonian
  $H^+_{\max}
  = H^+_{\max}(\Omega_1,\ldots,\Omega_n;\theta_1,\ldots,\theta_n)$
  has an infinitely degenerate zero mode if one of the following
  conditions is satisfied:

\medskip

\noindent (i) $\theta_1+\ldots+\theta_n<1$,

\medskip

\noindent (ii) $\displaystyle\frac{\theta_1}{\omega_1}
    +\ldots+\frac{\theta_n}{\omega_n}>\frac{1}{\omega_1}
    +\ldots+\frac{1}{\omega_n}
    -\frac{1}{\min\limits_{j}\,\omega_j}$\,,

\medskip

\noindent (iii) $n=2$.
\end{thm}

\begin{proof}
  (i) In this case one can choose numbers $p_j>1$ so that
  $p_j\theta_j<1$, $\forall\,j=1,\ldots,n$, and $\sum p_j^{-1}=1$
  (e.g., $p_j^{-1}=\theta_j/(\theta_1+\ldots+\theta_n)$).
  Let us consider the functions
  $$
  g_j(z,\bar{z})
  = \frac{\sin\!\big(\alpha_j(z-\kappa_j)\big)}{z-\kappa_j}
  \left|\sin\!\left(\frac{\pi(z-\kappa_j)}{\omega_j}
    \right)\right|^{-\theta_j} \,,
  $$
  where
  $$
  0<\alpha_j<\min\limits_{j}\frac{\pi}{\omega_j}\theta_j\,.
  $$
  Set $\psi=g_1\cdots g_n$. It suffices to show that $\psi$ is
  square integrable. From the Jensen's inequality it follows that
  \begin{equation}
    \label{6.4.1}
    |\psi|^2\le \frac{|g_1|^{2p_1}}{p_1}+\ldots+\frac{|g_n|^{2p_n}}{p_n}\,.
  \end{equation}
  Recalling that $p_j\theta_j<1$, $p_j>1$, and repeating the
  considerations from the proof of Theorem~\ref{orig:thm-6.3} one can
  show that each summand on the RHS of formula (\ref{6.4.1}) is
  integrable.

  Let us now discuss condition (ii). One can assume that
  $\min_{j}\omega_j=\omega_1$ and $\kappa_1=0$. We shall consider a
  function $\psi$ of the form
  $$
  \psi(z,\bar{z}) = \frac{\sin(\alpha{}z)}{z}\,
  g_1(z,\bar{z})\cdots g_n(z,\bar{z})
  $$
  where
  \begin{equation}
    \label{6.4.2}
    g_1(z,\bar z)= \left|\sin\!\left(\frac{\pi z}{\omega_1}
      \right)\right|^{-\theta_1}\,,
  \end{equation}
  \begin{equation}
    \label{6.4.3}
    g_j(z,\bar z)=
    \left|\sin\!\left(\frac{\pi(z-\kappa_j)}{\omega_j}
      \right)\right|^{1-\theta_j}\,,
  \end{equation}
  for $j=2,\ldots,n$, and $\alpha$ obeys the condition
  \begin{equation}
    \label{6.4.4}
    0<\alpha<\theta :=
    \frac{\pi}{\omega_1}-\pi\sum_j\frac{1-\theta_j}{\omega_j}\,.
  \end{equation}
  Let $T$ be a strip parallel to the real line $\R$ and containing
  $\R$ in its interior. One again concludes from (\ref{6.4a}) that
  outside the strip $T$ it holds true that
  \begin{displaymath}
    |\psi(z,\bar z)| \leq
    c_1\,\frac{\exp\!\left((\alpha-\theta)|y|\right)}{|z|}\,.
  \end{displaymath}
  Furthermore, inside $T$ we can use the estimate
  \begin{displaymath}
    |\psi(z,\bar z)| \leq c_2\left|\frac{\sin(\alpha{}z)}{z}\right|
    |g_1(z,\bar z)|.
  \end{displaymath}
  Therefore one can use a similar reasoning as in the proof of
  Theorem~\ref{orig:thm-6.3} to show that $\psi\in L^2(\R^2)$.

  Finally, let us discuss condition (iii). If $\omega_1=\omega_2$ then
  one can refer to Theorem~\ref{orig:thm-6.3}. In the opposite case we
  shall assume that $\omega_1<\omega_2$. If
  $\theta_1+\theta_2<1$ then we apply condition (i) from the
  theorem. If not then we have
  $$
  \frac{\theta_1}{\omega_1}+\frac{\theta_2}{\omega_2}>
  \frac{\theta_1+\theta_2}{\omega_2} \ge \frac{1}{\omega_2}
  = \frac{1}{\omega_1}+\frac{1}{\omega_2}-\frac{1}{\min\,\omega_j}\,,
  $$
  and we can apply condition (ii).
\end{proof}

According to Lemma~\ref{orig:lem-3.4} we can reformulate the result
for the operator $H_{\max}^-$ as follows.

\begin{thm}\label{orig:thm-6.4a}
  Assume that all chains $\Omega_j$, $j=1,\ldots,n$, are contained in
  $\R$. Then the Hamiltonian
  $H^-_{\max}
  =H^-_{\max}(\Omega_1,\ldots,\Omega_n;\theta_1,\ldots,\theta_n)$
  has an infinitely degenerate zero mode if one of the following
  conditions is satisfied:

\medskip

\noindent (i) $\theta_1+\ldots+\theta_n>n-1$,

\medskip

\noindent (ii) $\displaystyle\frac{\theta_1}{\omega_1}
    +\ldots+\frac{\theta_n}{\omega_n}<
    \frac{1}{\min\limits_{j}\,\omega_j}$\,,
\medskip

\noindent (iii) $n=2$.
\end{thm}

\subsection{A lattice of Aharonov--Bohm fluxes}

Let us now consider the Hamiltonian $H^{\pm}_{\max}$ for a lattice
of Aharonov--Bohm solenoids $\Omega=K+\Lambda$ where $\Lambda$ is
the Bravais lattice of the crystallographic lattice $\Omega$ with
a basis $\{\omega_1,\omega_2\}$ (cf. Example~7 in
Section~\ref{sec:examples-orig:2}). To analyze this case we shall
use the Weierstrass $\sigma$-function
$\sigma(z)\equiv\sigma(z;\omega_1,\omega_2)$. Let us introduce,
following \cite{Per,Per1}, the modified Weierstrass
$\sigma$-function $\tilde\sigma(z)$,
\begin{equation}
  \tilde\sigma(z)=e^{-\nu z^2}\sigma(z)
\end{equation}
where
$$
\nu =\frac{i}{4S}(\eta_1\bar\omega_2-\eta_2\bar\omega_1)\,,\quad
\eta_j=2\,\zeta\!\left(\frac{\omega_j}{2}\right)\,,\quad
S=\Im(\overline{\omega_1}\omega_2)\,,
$$
and $\zeta(z)$ is the Weierstrass $\zeta$-function (cf.
Appendix~\ref{sec:weierstr-sigma-funct}). We shall need the following
lemma. The number $\mu$, occurring in the formulation of lemma and
depending on the lattice $\Lambda$, is defined by
\begin{equation}
  \mu=\frac{\pi}{2S}\,.
\end{equation}

\begin{lem}\label{orig:lem-6.3}
  Let $\alpha_j$, $j=1,\ldots,n$, be real numbers such that
  $0<\alpha_j<1$, let $\beta$ be an arbitrary real number, and let
  $a_j$, $j=1,\ldots,n$, be an arbitrary array of complex numbers such
  that among them there is no couple congruent modulo $\Lambda$. If
  the condition
  $$
  \beta<\mu\sum\limits_{j=1}^n\alpha_j\,
  $$
  is satisfied then
  $$
  \exp(\beta|z|^2)\prod\limits_{j=1}^n
  |\tilde\sigma(z-a_j)|^{-\alpha_j}\in L^2(\R^2)\,.
  $$
\end{lem}

\begin{proof}
  We shall consider a shifted elementary cell of the lattice
  $\Lambda$,
  $$
  L_\epsilon=\{(t_1+\epsilon)\omega_1+(t_2+\epsilon)\omega_2;
  \textrm{ }0\le t_1,t_2<1\}\,,
  $$
  where $\epsilon>0$ is chosen so that the interior of $L_\epsilon$
  contains exactly one zero for each of the functions
  $\tilde\sigma(z-a_j)$, and hence exactly one pole of the function
  $1/\tilde\sigma(z-a_j)$. But in that case,
  \begin{equation}
    \label{13.1}
    \int\limits_{L_\epsilon}\bigg|\prod\limits_{j=1}^n
    \tilde\sigma(z-a_j)\bigg|^{-2\alpha_j}\, dxdy<\infty\,.
  \end{equation}
  Let $\rho(z,\bar z)$ be a function defined by the formula
  \begin{equation}
    \label{eq:def-rho_zz}
    |\sigma(z)|^2=\exp(\nu z^2+\bar\nu\bar z^2+2\mu z\bar{z})
    \rho(z,\bar z)\,,
  \end{equation}
  From formula (\ref{eq:def-rho_zz}) we deduce that
  \begin{equation}
    \label{13.2}
    I := \iint\limits_{L_\epsilon}
    \prod\limits_{j=1}^n|\rho(z-a_j,\bar z-\bar a_j)|^{-\alpha_j}\,
    dxdy<\infty\,.
  \end{equation}
  Since the function $\rho(z,\bar z)$ is $\Lambda$-periodic and
  $|\tilde\sigma(z)|^2=\exp(2\mu|z|^2)\rho(z,\bar z)$ (see
  Lemma~\ref{orig:lem-9.2}) it holds true that
  \begin{eqnarray*}
    & & \iint\limits_{\R^2}\exp(2\beta|z|^2)\bigg|
    \prod\limits_{j=1}^n\tilde\sigma(z-a_j)\bigg|^{-2\alpha_j}
    \,dxdy \\
    & & \qquad =\, \sum\limits_{\lambda\in\Lambda}\,
    \iint\limits_{L_{\epsilon}+\lambda}
    \exp\!\left(2(\beta-\mu\sum\alpha_j)|z|^2\right)
    \bigg|\prod\limits_{j=1}^n
    \rho(z-a_j,\bar z-\bar a_j)\bigg|^{-\alpha_j}
    \,dxdy \\
    & & \qquad \le\, I\sum\limits_{\lambda\in\Lambda}\,
    \sup\Big\{\exp\!\left(2(\beta-\mu\sum\alpha_j)|z|^2\right);
    \textrm{ }z\in L_{\epsilon}+\lambda\Big\}<\infty\,.
    \textrm{ }\qed
  \end{eqnarray*}
  \renewcommand{\qed}{}
\end{proof}

\begin{thm}\label{orig:thm-6.5}
  Let $\Omega=K+\Lambda$ be a lattice of Aharonov--Bohm solenoids with
  an array of fluxes $\Theta=(\theta_\kappa)_{\kappa\in{}K}$,
  $0<\theta_\kappa<1$. Then each of the operators
  $H^{\pm}_{\max}(\Omega;\Theta)$ has an infinitely degenerate zero
  mode.
\end{thm}

\begin{proof}
  We shall confine ourselves to the case of the operator $H^+_{\max}$.
  Let us consider the function
  \begin{equation}
    \psi(z,\bar z)=f(z)\prod\limits_{\kappa\in  K}
    |\tilde\sigma(z-\kappa)|^{-\theta_\kappa}\,.
  \end{equation}
  According to Lemma~\ref{orig:lem-6.3}, $\psi\in L^2(\R^2)$ if $f$ is
  an arbitrary polynomial.
\end{proof}

\begin{rem}\label{orig:rem-6.1}
  Owing to Theorem~\ref{orig:thm-6.5}, we can describe an interesting
  example related to the question of absolutely continuous spectrum
  for the Pauli operator $H_{\max}^{\pm}(\ba)$ with a magnetic field
  $b=\dd_xa_y-\dd_ya_x$ which is supposed to be periodic with respect
  to a lattice
  $\Lambda=\{k_1\omega_1+k_2\omega_2;\textrm{ }k_1,\,k_2\in\Z\}$, with
  $S=\Im(\bar\omega_1\omega_2)>0$. If the vector potential $\ba$ is
  ``sufficiently regular'' and the flux of the field $b$ through the
  elementary cell equals zero then the spectrum of the operators
  $H_{\max}^{\pm}(\ba)$ is purely absolutely continuous (see
  \cite{BS1,BS2,BSS,KL,Mor,Sob} and others). The same result is true
  for Schr\"odinger operators with ``sufficiently regular'' periodic
  vector potentials in the space $L^2(\R^d)$ for any $d\ge 2$. In the
  case $d\ge3$, N.~D. Filonov described in \cite{Fil} an example
  showing that the assumptions on the vector potential stated in
  \cite{Sob} and other papers cannot be essentially weakened.
  Theorem~\ref{orig:thm-6.5} shows that two-dimensional Pauli
  operators with a singular two-periodic magnetic field may have
  (infinitely degenerate) eigenvalues. In more detail, let us take,
  for example, a set $K$ containing two elements,
  $K=\{\kappa_1,\kappa_2\}$, and suppose that
  $\theta\equiv\theta_{\kappa_1}=-\theta_{\kappa_2}\in\,]0,1[$. Then
  by Theorem~\ref{orig:thm-6.5} the both operators
  $H_{\max}^{\pm}(\ba)$ have an eigenvalue, namely the number zero.
  According to Example~7 from Section~\ref{sec:examples-orig:2}, the
  corresponding vector potential $\ba$ reads
  $a_x=\theta\,\Im(\zeta(z-\kappa_1)-\zeta(z-\kappa_2))$,
  $a_y=\theta\,\Re(\zeta(z-\kappa_1)-\zeta(z-\kappa_2))$. Owing to
  quasi-periodicity of the Weierstrass function $\zeta(z)$,
  $\zeta(z+\omega_j)=\zeta(z)+\eta_j$ where
  $\eta_j=2\,\zeta(\omega_j/2)$, the vector potential $\ba$ is
  $\Lambda$-periodic.
\end{rem}

Now we state an analog of Theorem~\ref{orig:thm-6.3} for lattices of
Aharonov--Bohm fluxes. In view of the example described in Remark
\ref{orig:rem-9.1}, we cannot here repeat the arguments from the proof
of Theorem~\ref{orig:thm-6.3} but the properties of the modified
Weierstrass $\sigma$-function simplify matters considerably.

\begin{thm}
  \label{orig:thm-New}
  Let a uniformly discrete set $\Omega$ be expressible as a disjoint
  union of a finite number of lattices $\Omega_1\,,\ldots,\Omega_n$,
  and let the lattice $\Omega_j=K_j+\Lambda_j$ carry Aharonov--Bohm
  fluxes $(\theta_\kappa)_{\kappa\in{}K_j}$ ($j=1\,,\ldots,\,n$,
  $0<\theta_\kappa<1$). Then the Hamiltonians $H_{\max}^{\pm}(\Omega)$
  have infinitely degenerate zero modes.
\end{thm}

\begin{proof}
  In the proof we shall consider only the operator $H_{\max}^+$.
  Without loss of generality we suppose that each $K_j$ is a
  singleton: $K_j=\{\kappa_j\}$, and we shall write $\theta_j$ instead
  of $\theta_{\kappa_j}$. By the hypothesis, there is a sufficiently
  small disk $D$ centered at $0$ such that for
  $\omega_1\,,\omega_2\in\Omega$, $\omega_1\ne\omega_2$, the sets
  $D+\omega_1$ and $D+\omega_2$ are disjoint.  Denote
  $L_j:=D+\kappa_j+\Lambda_j$\,, then for every $j$ there exists
  $c_j>0$ such that $|\tilde\sigma(z-\kappa_j)|^{-\theta_j}\le c_j$
  for $z\notin L_j$. It is clear that
  $$
  \prod\limits_{j=1}^n
  |\tilde\sigma(z-a_j)|^{-\theta_j}\le
  \sum\limits_{j=1}^n\,\big(\,\prod_{k\ne j}\,c_k \big)
  |\tilde\sigma(z-a_j)|^{-\theta_j}\,.
  $$
  Now we can refer to Lemma~\ref{orig:lem-6.3}.
\end{proof}

Analogs of Theorem~\ref{orig:thm-6.4} and
Theorem~\ref{orig:thm-6.4a} are also valid and can be proved by
the same method. In more detail, let $\Omega_j$, $j=1,\ldots,n$,
be mutually disjoint simple crystallographic lattices,
$\Omega_j=\kappa_j+\Lambda_j$ where $\kappa_j\in\C$,
$\Lambda_j=\{k_1\omega_1^{(j)}+k_2\omega_2^{(j)};\textrm{ }
k_1,\,k_2\in\Z\}$.  Furthermore,
$S_j=\Im(\bar{\omega}_1^{(j)}\omega_2^{(j)})$ designates the area
of the elementary cell of the Bravais lattice $\Lambda_j$.

\begin{thm}
  The Hamiltonian
  $H^+_{\max}=H^+_{\max}(\Omega_1,\ldots,\Omega_n;\theta_1,\ldots,\theta_n)$
  has an infinitely degenerate zero mode if one of the following
  conditions is satisfied:

  \medskip

  \noindent (i) $\theta_1+\ldots+\theta_n<1$,

  \medskip

  \noindent (ii) $\displaystyle\frac{\theta_1}{S_1}+\ldots+
  \frac{\theta_n}{S_n}> \frac{1}{S_1}+\ldots+
  \frac{1}{S_n}-\frac{1}{\min\limits_j\,S_j}$\,,

  \medskip

  \noindent (iii) $n=2$ and $S_1\ne S_2$\,.
\end{thm}

\begin{thm}
  The Hamiltonian
  $H^-_{\max}
  =H^-_{\max}(\Omega_1,\ldots,\Omega_n;\theta_1,\ldots,\theta_n)$
  has an infinitely degenerate zero mode if one of the following
  conditions is satisfied:

  \medskip

  \noindent (i) $\theta_1+\ldots+\theta_n>n-1$,

  \medskip

  \noindent (ii) $\displaystyle\frac{\theta_1}{S_1}+\ldots+
  \frac{\theta_n}{S_n}< \frac{1}{\min\limits_j\,S_j}$\,,

  \medskip

  \noindent (iii) $n=2$ and $S_1\ne S_2$\,.
\end{thm}

\subsection{Superposition of a homogeneous magnetic field
  with a field corresponding to Aharonov--Bohm solenoids}

Here we consider a perturbation of a homogeneous magnetic field by the
field corresponding to a system of Aharonov--Bohm solenoids, i.e., we
consider a vector potential $\ba$ of the form $\ba=\ba_0+\ba_{AB}$
where $\ba_0$ is the vector potential of a homogeneous magnetic field
$b_0=2\pi\xi_0$ with a flux density $\xi_0$, $\ba_0=\pi\xi_0 (-y,x)$,
and $\ba_{AB}$ is the vector potential of a system of Aharonov--Bohm
fluxes. We shall suppose that the potential $\ba_{AB}$ is of finite
type. In that case we have a finite family of mutually disjoint
discrete subsets in the complex plane, $\Omega_1,\ldots,\Omega_n$, and
in each point of the set $\Omega_j$ ($j=1,\ldots,n$) there is a flux
of magnitude $\theta_j$ ($0<\theta_j<1$) intersecting the plane.

Suppose for definiteness that $b_0>0$. Then the operator
$H_{\max}^+(\ba_0)$ has an infinitely degenerate zero mode (the lowest
Landau level shifted by the value $-b_0$) while the ground state of
the operator $H_{\max}^-(\ba_0)$ is strictly positive (this is the
lowest Landau level shifted by the value $b_0$). Thus the latter
operator has no zero mode. Intuitively, the results proved below in
Theorems~\ref{orig:thm-6.7} and \ref{orig:thm-6.8} mean that if the
set $\Omega=\Omega_1\cup\ldots\cup\Omega_n$ has a finite density then
a superposition with the potential $\ba_{AB}$ does not remove the zero
mode from the spectrum of the operator $H_{\max}^+$, and a zero mode
cannot occur in the spectrum of the operator $H_{\max}^-$ provided
$b_0$ is {\em sufficiently large}.  Moreover, if this set has zero
density then the same statement about zero modes of the operators
$H_{\max}^+$ and $H_{\max}^-$ is true for {\em any} $b_0>0$. In the
case when $\Omega$ is a lattice, a superposition with the potential
$\ba_{AB}$ does not remove the zero mode from the spectrum of
$H_{\max}^+$ for {\em any} $b_0>0$ but a zero mode may occur in the
spectrum of $H_{\max}^-$ for {\em particular values} of fluxes
$\theta_j$. An attentive reader can effortlessly guess what happens
for $b_0<0$.

Let $\alpha$ be an arbitrary positive number. For any $r>0$ we denote
\begin{equation}
  \label{6.7.1}
  S(r)=\sum\limits_{\omega\in\Omega, 0<|\omega|\le r} \omega^{-\alpha}\,,
\end{equation}
\begin{equation}
  \label{6.7.2}
  T(r)=\sum\limits_{\omega\in\Omega, 0<|\omega|\le r} |\omega|^{-\alpha}\,,
\end{equation}
\begin{equation}
  \label{6.7.3}
  n(r)=\#\{\omega\in\Omega;\, |\omega|\le r\}\,.
\end{equation}

\begin{thm}\label{orig:thm-6.7}
  Suppose that $\ba=\ba_0+\ba_{AB}$ and that the Aharonov--Bohm vector
  potential $\ba_{AB}$ is of finite type. Let the following conditions
  be satisfied: $(a)$ for any $\alpha>2$, the sums $T(r)$ are
  uniformly bounded, $(b)$ $n(r)=O(r^2)$, $(c)$ the sums $S(r)$ are
  uniformly bounded for $\alpha=2$. Then, for sufficiently large
  $b_0>0$, the Hamiltonian $H_{\max}^+(\ba)$ has an infinitely
  degenerate zero mode and $H_{\max}^-(\ba)$ has no zero mode.

  If for $\alpha=2$ the sums $T(r)$ are uniformly bounded then, for
  any $b_0>0$, the Hamiltonian $H_{\max}^+(\ba)$ has an infinitely
  degenerate zero mode and $H_{\max}^-(\ba)$ has no zero mode.

  In the case when $b_0<0$ the same claim remains true when
  interchanging the role of $H_{\max}^+(\ba)$ and $H_{\max}^-(\ba)$.
\end{thm}

\begin{proof}
  Let us consider the operator $H_{\max}^+(\ba)$. In view of
  Theorem~\ref{orig:thm-5.1}, we can assume that its zero mode, if
  any, has the form
  \begin{equation}
    \label{6.7.4}
    \psi(z,\bar z)=f(z)\exp\!\left(-\frac{1}{2}\pi\xi_0|z|^2\right)
    \prod_{j=1}^n|W_j|^{-\theta_j}
  \end{equation}
  where $W_j(z)=W_{\Omega_j}(z)$ is the Weierstrass canonical product
  for the set $\Omega_j$ (see Appendix~\ref{sec:append-analyt-fce})
  and $f(z)$ is a nonzero entire function (cf.
  Lemma~\ref{orig:lem-6.2}).

  Let assumptions (a), (b), (c) be satisfied. Then, according to the
  Borel theorem and to the Lindel\"of theorem (cf.
  Theorems~\ref{orig:thm-9.1} and \ref{orig:thm-9.4} in Appendix),
  every function $W_j(z)$ has order 2 and a finite type, i.e.,
  $|W_j(z)|\le{}a_j\exp(c_j|z|^2)$ with some constants $a_j,\,c_j>0$.
  It follows that for
  \begin{equation}\label{eq:b0bound}
    \frac{b_0}{4} > c_1(1-\theta_1)+\ldots+c_n(1-\theta_n)
  \end{equation}
  the function (\ref{6.7.4}) is square integrable if we set
  $f(z)=p(z)\prod_{j=1}^nW_j(z)$ where $p(z)$ is an arbitrary
  polynomial.

  If the functions $T(r)$ are uniformly bounded for $\alpha=2$ then
  $p_\Omega\le1$ (see (\ref{3.7})) and, according to
  Theorem~\ref{orig:thm-9.5} from Appendix, the functions $W_j(z)$ are
  of minimal type and so the constants $c_j$ can be chosen arbitrarily
  small. Consequently, the restriction on the field $b_0>0$ is not
  necessary anymore.

  In the case of the operator $H_{\max}^-(\ba)$ we have to discuss the
  function
  \begin{equation}
    \label{6.7.5}
    \psi(z,\bar z)=f(z)\exp\!\left(\frac{1}{2}\pi\xi_0|z|^2\right)
    \prod_{j=1}^n|W_j|^{\theta_j-1}\,.
  \end{equation}
  If assumptions (a), (b), (c) are satisfied then
  $|W_j(z)|^{\theta_j-1}\ge a_j\exp(c_j(\theta_j-1)|z|^2)$
  with some constants $a_j,\,c_j>0$. Consequently, for $b_0$ obeying
  (\ref{eq:b0bound}), $R>0$ sufficiently large and for some $c>0$ we
  have the inequality $|\psi(z)|^2\ge{}c|f(z)|^2$ if $|z|\ge R$. But
  then, as one deduces from Lemma~\ref{orig:lem-6.1}, $\psi$ is not
  square integrable, hence $H_{\max}^-(\ba)$ has no zero modes.

  Obviously, changing the sign at $b_0$ means that $H_{\max}^+(\ba)$
  and $H_{\max}^-(\ba)$ interchange their roles in the above
  considerations.
\end{proof}

\begin{rem}
  If conditions (a), (b), (c) from Theorem~\ref{orig:thm-6.7} are
  satisfied then $\Omega$ has finite density, i.e.,
  $\limsup\limits_{r\to \infty}\,n(r)/r^2<\infty$. If, in addition,
  the sums $T(r)$ are uniformly bounded for $\alpha=2$ then the
  density of the set $\Omega$ is zero (see inequalities (\ref{2A.4})).
\end{rem}

\begin{rem}
  All assumptions of Theorem~\ref{orig:thm-6.7} are fulfilled if every
  set $\Omega_j$ is either finite or a union of chains.
\end{rem}

Let now $\Omega$ be a lattice. Using the above introduced notation we
write $\Omega_j=\kappa_j+\Lambda$ where
$\Lambda=\{k_1\omega_1+k_2\omega_2;\textrm{ }k_1,k_2\in\Z\}$. Suppose
that $S=\Im(\bar \omega_1\omega_2)>0$ ($S$ is the area of an
elementary cell in the lattice $\Lambda$). Let $\eta_0=\xi_0S$
designate the flux of the homogeneous component of the field through
the elementary cell of the lattice $\Lambda$.

\begin{thm}\label{orig:thm-6.8}
  Suppose that $\ba=\ba_0+\ba_{AB}$ and that $\Omega$ is a lattice.
  Let $b_0>0$. Then the Hamiltonian $H_{\max}^+(\ba)$ has an
  infinitely degenerate zero mode. The inequality
  $$
  \eta_0+\sum_{j=1}^n\theta_j < n
  $$
  is a sufficient and necessary condition for $H_{\max}^-(\ba)$ to
  have a zero mode, and if it is fulfilled then the zero mode is
  infinitely degenerate.

  Let $b_0<0$. Then the Hamiltonian $H_{\max}^-(\ba)$ has an
  infinitely degenerate zero mode. The inequality
  $$
  |\eta_0|<\sum_{j=1}^n\theta_j
  $$
  is a sufficient and necessary condition for $H_{\max}^+(\ba)$ to
  have a zero mode, and if it is fulfilled then the zero mode is
  infinitely degenerate.
\end{thm}

\begin{proof}
  We shall start the proof from the operator $H^+_{\max}$. In analogy
  with the proof of Theorem~\ref{orig:thm-6.5} we consider the
  function
  \begin{equation}
    \psi(z,\bar z)=f(z)\exp\!\left(-\frac{\pi\eta_0}{2S}|z|^2\right)
    \prod_{j=1}^n|\tilde\sigma(z-\kappa_j)|^{-\theta_j}\,
  \end{equation}
  (recall that $\eta_0=\xi_0S$). From Lemma~\ref{orig:lem-6.3} we
  immediately deduce that for an arbitrary polynomial $f$ it holds
  true that $\psi\in L^2(\R^2)$, hence $\psi$ is an infinitely
  degenerated zero mode.

  Let us now turn to the operator $H^-_{\max}$. According to
  Theorem~\ref{orig:thm-4.1} and Lemma~\ref{orig:lem-6.2}, the ground
  state of the operator $H_{\max}^-$, if any, has the form
  \begin{equation}
    \label{14.3}
    \psi(z,\bar z)=f(\bar z)
    \exp\!\left(\frac{\pi\eta_0}{2S}|z|^2\right)
    \prod_{j=1}^n
    |\tilde\sigma(z-\kappa_j)|^{\theta_j-1}\,
  \end{equation}
  where $f(z)$ is an entire function. Using formula (\ref{0.2}) from
  Appendix~\ref{sec:weierstr-sigma-funct} (where $\mu=\pi/2S$) we get
  \begin{eqnarray}
    |\psi(z,\bar z)|^2 &=& |f(\bar z)|^2
    \exp\!\left(\frac{\pi\eta_0}{S}|z|^2\right)\prod_{j=1}^n
    \exp\!\left(\frac{\pi}{S}(\theta_j-1)|z|^2\right) \nonumber\\
    & &\times\, \prod_{j=1}^n
    |\rho(z-\kappa_j,\bar z-\bar\kappa_j)|^{2(\theta_j-1)}
    \label{14.3b}\\
    &=& |f(\bar z)|^2\exp(c|z|^2)\prod_{j=1}^n|
    \rho(z-\kappa_j,\bar z-\bar\kappa_j)|^{2(\theta_j-1)}
    \nonumber
  \end{eqnarray}
  where
  $$
  c=\frac{\pi}{S}\left(\eta_0+\sum_{j=1}^n
    (\theta_j-1)\right).
  $$
  The condition $\eta_0+\sum_{j=1}^n\theta_j<n$ is
  equivalent to the condition $c<0$. But in that case the membership
  $\psi\in L^2(\R^2)$ can be proved as in Lemma~\ref{orig:lem-6.3}.

  Conversely, assume that $c\ge0$ and that there exists a non-zero
  entire function $f(z)$ such that $\psi\in L^2(\R^2)$. Then from
  (\ref{14.3b}) we derive that
  $|f(\bar{z})|^2\le{}c_1|\psi(z,\bar{z})|^2$ with some constant
  $c_1$.  Consequently, $f\in L^2(\R^2)$ which contradicts
  Lemma~\ref{orig:lem-6.1}.

  In the case $b_0<0$ one can either repeat the above reasoning or
  apply Corollary~\ref{orig:coroflem-3.4} while noticing that
  $\{-x\}=1-\{x\}$ for any $x\in\R$ which is not an integer.
\end{proof}

\begin{rem}
  Similarly to the case $b_0=0$ (cf. Remark~\ref{orig:rem-6.1}), both
  operators $H^{\pm}_{\max}(\ba)$ may have localized states also when
  the total flux through the elementary cell is zero.  Suppose, for
  example, that $b_0>0$, $0<\theta_1<1$, $0<\eta_0+\theta_1<1$ and
  $\theta_2=1-\eta_0-\theta_1$. Then $\eta_0+\theta_1+\theta_2=1<2$
  and the assumption of the theorem is satisfied.
\end{rem}

\begin{rem}
  Theorem~\ref{orig:thm-6.8} shows that, for $b_0>0$, an oscillation
  of the type ``localization--delocalization'' occurs after adding an
  Aharonov--Bohm flux to a system with the Hamiltonian $H^-_{\max}$.
\end{rem}

\section{Conservation of zero modes under translations and additions
  of Aharonov--Bohm solenoids. Irregular Aharonov--Bohm systems}
\label{sec:conserv-orig:7}

\subsection{Translation and addition of finitely many Aharonov--Bohm
  solenoids}

Up to now we have investigated zero modes of regular Aharonov--Bohm
systems (in the sense of the definition given at the end of
Section~\ref{sec:examples-orig:2}), with Theorem~\ref{orig:thm-6.7}
representing the only exception. The proof of this theorem suggests
that one should expect zero modes also in the case when the
homogeneous component of the magnetic field is absent provided the
perturbation corresponds to a (in general, irregular) ``sufficiently
scarce'' system. Further we shall consider such scarce perturbations
applied to systems of chains or lattices of Aharonov--Bohm solenoids.
Before addressing this question we shall prove that the zero mode of
the Hamiltonian corresponding to a system of solenoids of finite type
does not disappear if a finite number of solenoids are moved or if one
joins a finite number of solenoids to the system.

In the followings theorems, $\ba$ designates a potential of finite
type corresponding to a system of Aharonov--Bohm solenoids which is
determined by an array of mutually disjoint discrete sets,
$\Omega_1,\ldots,\Omega_n$, and by an array of fluxes,
$\theta_1,\ldots,\theta_n$ ($0<\theta_j<1$). $\Omega$ designates the
union $\Omega=\Omega_1\cup\ldots\cup\Omega_n$.

\begin{thm}\label{orig:thm-7.1}
  In addition to the above introduced notation let
  $K_j=\{\omega_{1j},\ldots,\omega_{n_j,j}\}$ be a finite subset of
  $\Omega_j$ and let $K'_j=\{\omega'_{1j},\ldots,\omega'_{n_j,j}\}$ be
  a finite subset of $\C$ such that the sets
  $\Omega'_j=(\Omega_j\setminus{}K_j)\cup{}K'_j$, $j=1,\ldots,n$, are
  mutually disjoint. If the operator $H_{\max}^{\pm}(\ba)$ has a zero
  mode then the operator $H_{\max}^{\pm}(\ba')$ determined by the
  array $(\Omega'_1,\ldots,\Omega'_n;\,\theta_1,\ldots,\theta_n)$ has
  also a zero mode with the same multiplicity as that of the zero mode
  for the operator $H_{\max}^{\pm}(\ba)$.
\end{thm}

\begin{proof}
  We shall confine ourselves to the discussion of the operator
  $H_{\max}^+(\ba)$. The zero modes of this operator can be written in
  the form
  $$
  \psi(z,\bar z)=f(z)\prod\limits_{j=1}^n\,|W_j(z)|^{-\theta_j}
  $$
  where $f$ is an entire function and $W_j(z)=W_{\Omega_j}(z)$ is
  the Weierstrass canonical product for $\Omega_j$. Then the function
  $$
  \tilde\psi(z,\bar z)=f(z)\prod\limits_{j=1}^n\,
  \left(|W_j(z)|^{-\theta_j} \prod\limits_{k=1}^{n_j}
    \bigg|\frac{z-\omega_{kj}}{z-\omega'_{kj}}
    \bigg|^{\theta_j}\right)
  $$
  represents a zero mode of $H_{\max}^+(\ba')$. One has only to
  verify that $\tilde\psi\in L^2(\R^2)$. It is actually so because the
  additional singularities at the points $\omega'_{kj}$ are square
  integrable and outside a compact set $\tilde\psi$ differs from
  $\psi$ by a bounded factor.

  This argument clearly shows that the multiplicity of the zero mode
  for the operator $H_{\max}^{\pm}(\ba')$ is not smaller than the
  multiplicity of the zero mode for the operator
  $H_{\max}^{\pm}(\ba)$.  Since the operators play an equivalent role
  in the assumptions the converse is also true.
\end{proof}

\begin{thm}
  Assume that additionally to the considered system of solenoids there
  are given a finite set
  $\Omega'=\{\omega'_1,\ldots,\omega'_m\}\subset\C$ not intersecting
  $\Omega$ and a corresponding family of fluxes
  $\{\theta'_1,\ldots,\theta'_m\}$ ($0<\theta'_j<1$). Let $\ba'$ be
  the vector potential determined by the array of sets
  $\Omega_1,\ldots,\Omega_n,\Omega'$, and by the array of fluxes
  $\theta_1,\ldots,\theta_n,\theta'_1,\ldots,\theta'_m$. If the
  operator $H_{\max}^{\pm}(\ba)$ has a zero mode then the operator
  $H_{\max}^{\pm}(\ba')$ also has a zero mode whose multiplicity is
  not smaller than the multiplicity of the zero mode for the operator
  $H_{\max}^{\pm}(\ba)$.
\end{thm}

\begin{proof}
  We shall confine ourselves to the discussion of the operator
  $H_{\max}^+(\ba)$. The zero modes of this operator can be written in
  the form
  $$
  \psi(z,\bar z)=f(z)\prod\limits_{j=1}^n\,|W_j(z)|^{-\theta_j}
  $$
  where $f$ is an entire function and $W_j(z)=W_{\Omega_j}(z)$ is
  the Weierstrass canonical product for $\Omega_j$. It turns out that
  the function
  $$
  \tilde\psi(z,\bar z)=f(z)\prod\limits_{j=1}^n\,
  |W_j(z)|^{-\theta_j}
  \prod\limits_{k=1}^{m}\left|z-\omega'_k\right|^{-\theta'_k}\,.
  $$
  is a zero mode of $H_{\max}^+(\ba')$. One has only to verify that
  $\psi\in L^2(\R^2)$. This again follows from the fact that the
  singularities at the points $\omega'_k$ are square integrable and
  the function $\tilde\psi$ differs from $\psi$ by a bounded factor
  outside a compact set.
\end{proof}

\subsection{Additional notation}

Up to the end of the current section, $\Omega_j$ ($j=1,\ldots,n$)
designates either an array of mutually disjoint chains or an array of
mutually disjoint lattices, $\Omega_j=K_j+\Lambda_j$ where $\Lambda_j$
is a Bravais lattice of rank 1 or 2 and
$K_j=\{\kappa_{1j},\ldots,\kappa_{m_j,j}\}$ is a finite set.  To each
set $\Omega_j$ we relate an array of Aharonov--Bohm fluxes
$\Theta_j=(\theta_{kj})_{1\le k\le m_j}$. By $\Omega$ we denote the
union $\Omega=\Omega_1\cup\ldots\cup\Omega_n$. In addition, we shall
consider another array of discrete subsets in the plane,
$\Omega'_1,\ldots,\Omega'_m$, whose members are mutually disjoint as
well as disjoint with the sets $\Omega_1,\ldots,\Omega_n$, and we
relate to these additional sets an array of fluxes
$\Theta'=(\theta'_j)_{1\le j\le m}$. Finally, we consider a discrete
set $\Omega'_0\subset\Omega$ whose points are supposed to be removed
from $\Omega$. Set
$\Omega'=\Omega'_0\cup\Omega'_1\cup\ldots\cup\Omega'_m$. Furthermore,
$W_j(z)$ is the Weierstrass canonical product for the set $\Omega'_j$,
$V_{kj}(z)$ is the Weierstrass canonical product for the set
$\Omega'_0\cap(\kappa_{kj}+\Lambda_j)$. By $\tau'$ we denote the
convergence exponent of the set $\Omega'$ and by $p'$ its genus. The
symbol $n'(r)$ designates the number of points of the set $\Omega'$
contained in the disc $|z|\le r$. The symbol $\ba'$ designates the
vector potential for the {\em perturbed} system of solenoids
determined by the discrete sets
$\Omega_1\setminus\Omega'_0,\ldots,\Omega_n\setminus\Omega'_0,
\Omega'_1\cup\ldots\cup\Omega'_m$, and by the array of fluxes obtained
by concatenating the arrays $\Theta_1,\ldots,\Theta_n$ and $\Theta'$.
Let us note that we still assume that $0<\theta_j<1$, $0<\theta'_j<1$.

Thus the set $\Omega'$ representing the total perturbation of the
original set $\Omega$ need not be finite nor regular. On the other
hand, we shall always suppose that the set $\Omega'$ is sufficiently
scarce by imposing restrictive assumptions on its genus $p'$.

\subsection{Addition of solenoids to a union of Aharonov--Bohm chains}

In Theorem~\ref{orig:thm-6.3} it has been shown, roughly, that if
$\Omega$ is a finite union of chains then the Hamiltonians
$H_{\max}^{\pm}(\Omega)$ have infinitely many zero modes. Below we
show that this property survives provided the perturbation $\Omega'$
is sufficiently scarce and at least too chains are not parallel.

\begin{thm}\label{orig:thm-7.3}
  Let $\Omega_1,\ldots,\Omega_n$ be chains whose union is a uniformly
  discrete set. Suppose that among the chains there are at least two
  which are not parallel. Furthermore, suppose that the genus of
  $\Omega'$ fulfills $p'=0$.  Then the Hamiltonians
  $H_{\max}^{\pm}(\ba')$ have infinitely degenerate zero modes.
\end{thm}

\begin{proof}
  In virtue of Lemma~\ref{orig:lem-9.1} one can assume that every
  chain $\Omega_j=K_j+\Lambda_j$, with $\Lambda_j=\{k\omega_j;\textrm{
  }k\in\Z\}$, $\omega_j\ne0$, is contained in a line $L_j$ and that
  different chains are contained in different lines. We shall suppose,
  without loss of generality, that $L_1$ and $L_2$ are not parallel
  and that $L_1$ coincides with the real line $\R$, with
  $0\in\Omega_1$.

  Let us consider a function $\psi$ of the form
  \begin{equation}
    \psi(z,\bar z)=f(z)\prod_{j=1}^ng_j(z,\bar z)
    \prod_{k=1}^m|W_k(z)|^{1-\theta'_k}
  \end{equation}
  where
  \begin{equation}
    g_j(z,\bar z)=\prod_{k=1}^{m_j}\left|\sin\!\left(
        \frac{\pi(z-\kappa_{kj})}{\omega_j}\right)\right|^{-\theta_{kj}}
    |V_{kj}(z)|^{\theta_{kj}}\,.
  \end{equation}
  To prove the theorem it suffices to find an infinite, linearly
  independent family of entire function $f(z)$ such that
  $\psi\in{}L^2(\R^2)$.

  Let us show that the functions
  \begin{equation}
    f(z)=\frac{\sin(\alpha z)}{z}
  \end{equation}
  with sufficiently small $\alpha>0$ suit this condition. To this end
  we shall need the following lemma.  \renewcommand{\qed}{}
\end{proof}

\begin{lem}\label{orig:lem-7.1}
  Assume that $\alpha_1,\alpha_2,a_1,a_2\in\C$ fulfill
  $\alpha_1\alpha_2\ne0$, $\alpha_1/\alpha_2\notin\R$. Then for every
  $\epsilon>0$ there exist constants
  $\tilde{c},c_1,c_2,\gamma_1,\gamma_2>0$ such that the inequality
  \begin{equation}
    \label{72.4}
    c_1e^{\gamma_1|z|}\le|\sin\!\big(\alpha_1(z-a_1)\big)
    \sin\!\big(\alpha_2(z-a_2)\big)|\le c_2e^{\gamma_2|z|}
  \end{equation}
  holds true whenever $|z|\ge\tilde c$ and the distance from $z$ to
  the lines $L_1=\alpha_1^{-1}\R+a_1$ and $L_2=\alpha_2^{-1}\R+a_2$ is
  greater than $\epsilon$.
\end{lem}

\begin{proof}[Proof of Lemma~\ref{orig:lem-7.1}]
  Let $L$ be a line written in the form $L=\alpha^{-1}\R+a$ where
  $\alpha,a\in\C$, $\alpha\ne 0$. Then inequality (\ref{6.4a}) implies
  that
  \begin{equation}
    \label{72.5}
    \frac{e^{|\alpha|d}-1}{2}
    \le|\sin\!\big(\alpha(z-a)\big)|\le e^{|\alpha|d}
  \end{equation}
  where $d$ is the distance from the point $z$ to the line $L$.
  Actually, set $\alpha=|\alpha|e^{i\phi}$. Then
  $|\Im\alpha(z-a)|=|\alpha|d$ where $d$ is the distance from
  $e^{i\phi}(z-a)$ to $\R$. But, at the same time, $d$ is the distance
  from $z$ to $e^{-i\phi}\R+a=\alpha^{-1}\R+a$. From (\ref{72.5}) we
  deduce that for every $\epsilon>0$ one can find a constant $c>0$
  such that
  \begin{equation}
    \label{72.6}
    |\sin\!\big(\alpha(z-a)\big)|\ge ce^{|\alpha|d}
  \end{equation}
  whenever the distance $d$ from $z$ to $L$ is greater than
  $\epsilon$. Actually, it suffices to choose
  $c=(1-e^{-|\alpha|\epsilon})/2$.

  Let us now reconsider the lines $L_1$ and $L_2$ from the lemma.
  Since $\alpha_1/\alpha_2\notin\R$ the lines intersect each other in
  a point $v\in\C$. Let us denote by $\phi$ the angle between the
  lines $L_1$ and $L_2$ ($0<\phi<\pi$), and by $\theta$ the angle
  between the vector $z-v$ and the line $L_1$. Then the distances
  $d_1$ and $d_2$ from the point $z$ to the lines $L_1$ and $L_2$ are
  respectively equal
  $$
  d_1=|z-v||\sin(\theta)|\,,\quad d_2=|z-v||\sin\,(\theta-\phi)|\,.
  $$
  From (\ref{72.5}) we get
  $$
  |\sin\!\big(\alpha_1(z-a_1)\big) \sin\!\big(\alpha_2(z-a_2)\big)|
  \le e^{|\alpha_1|d_1+|\alpha_2|d_2} \le
  e^{\max(|\alpha_1|,|\alpha_2|)(d_1+d_2)}\,.
  $$
  Notice that $d_1+d_2\le2|z-v|\le 2|z|+2|v|$, hence
  $$
  |\sin\!\big(\alpha_1(z-a_1)\big)
  \sin\!\big(\alpha_2(z-a_2)\big)|\le c_2e^{\gamma_2|z|}
  $$
  where
  $$
  \gamma_2=2\max(|\alpha_1|,|\alpha_2|)\,,\textrm{ }
  c_2=e^{|v|\gamma_2}\,,
  $$
  (this is true for any $z\in\C$).

  Using (\ref{72.6}) we can relate to every $\epsilon>0$ a constant
  $c>0$ such that if $d_1,d_2>\epsilon$ then
  $$
  |\sin\!\big(\alpha(z-a_1)\big)\sin\!\big(\alpha(z-a_2)\big)|
  \ge c^2e^{\min(|\alpha_1|,|\alpha_2|)(d_1+d_2)}\,.
  $$
  On the other hand,
  \begin{eqnarray*}
    d_1+d_2 &\ge& |z-v|(\sin^2\theta+\sin^2(\theta-\phi))
    = |z-v|(1-\cos\phi\cos(\phi-2\theta)) \\
    &\ge& |z-v|(1-|\cos\phi|)
    \ge |z|(1-|\cos\phi|)-|v|(1-|\cos\phi|)\,.
  \end{eqnarray*}
  From here we deduce that the inequality
  $$
  |\sin\!\big(\alpha(z-a_1)\big)
  \sin\!\big(\alpha(z-a_2)\big)|\ge c_1e^{\gamma_1|z|}
  $$
  holds true for
  $$
  c_1=c^2e^{-\min(|\alpha_1|,|\alpha_2|)|v|
    (1-|\cos\phi|)}\,,\textrm{ }
  \gamma_1=\min(|\alpha_1|,|\alpha_2|)(1-|\cos\phi|)\,.
  $$
  This proves the lemma.
\end{proof}

\begin{proof}[Proof of Theorem~\ref{orig:thm-7.3} (continued)]
  Let us return to the proof of Theorem~\ref{orig:thm-7.3}. From the
  assumptions of the theorem ($p'=0$) it follows that the functions
  $V_{kj}(z)$ and $W_j(z)$ are of growth $(1,0)$ (see
  Appendix~\ref{sec:append-analyt-fce}). This implies that for every
  $\epsilon>0$ there exists a constant $c_\epsilon>0$ such that
  \begin{equation}
    \label{72.7}
    \prod_{l=1}^m|W_l(z)|^{1-\theta'_l}\prod_{j=1}^{n}\prod_{k=1}^{m_j}
    |V_{kj}(z)|^{\theta_{kj}}\le c_\epsilon\exp(\epsilon|z|)\,.
  \end{equation}

  Let $P_j$ be a strip with border lines parallel to $L_j$ and
  containing $L_j$ in its interior, and let $Q$ be a sufficiently
  large disk centered at 0. In virtue of Lemma~\ref{orig:lem-7.1}, the
  disk $Q$ can be chosen so that, for $z\notin{}Q\cup P_1\cup P_2$,
  the inequality
  \begin{equation}
    \label{72.8}
    |\psi(z,\bar z)|\le c_1|f(z)|
    |\tilde g_1^B(z,\bar z)||\tilde g_2^B(z,\bar z)|
    \prod_{j=3}^{n}|\tilde g_j^A(z,\bar z)|
  \end{equation}
  holds true with some constant $c_1>0$.  Here we have set
  $$
  \tilde g_j^A(z,\bar z)=
  \prod_{k=1}^{m_j}\left|\sin\!\left(\frac{\pi(z-\kappa_{kj})}{\omega_j}
    \right)\right|^{-\theta_{kj}}
  $$
  and
  $$
  \tilde g_j^B(z,\bar z)=\left|\sin\!\left(
      \frac{\pi(z-\kappa_{1j})}{\omega_j}\right)\right|^{-\beta_j}
  \prod_{k=2}^{m_j}\left|\sin\!\left(
      \frac{\pi(z-\kappa_{jk})}{\omega_j}
    \right)\right|^{-\theta_{kj}},
  $$
  with $0<\beta_j<\theta_{1j}$.

  On the other hand, one can choose $Q$ so that, for
  $z\in{}P_2\setminus{}Q$, we have the inequality $|z|\le c'd_1$ where
  $d_1$ is the distance from $z$ to the line $L_1$, and $c'>0$ does
  not depend on $z$. Then from (\ref{72.6}) we deduce that $Q$ can be
  replaced by a larger disk such that the inequality
  \begin{equation}
    \label{72.8b}
    |\psi(z,\bar z)|\le c'_1|f(z)||\tilde g_1^B(z,\bar z)|
    \prod_{j=2}^{n}|\tilde g_j^A(z,\bar z)|\,,
  \end{equation}
  with a constant $c'_1>0$, holds true for $z\in P_2\setminus Q$. An
  analogous assertion is true when interchanging the strips $P_1$ and
  $P_2$.

  Finally, for any choice of the disk $Q$, the inequality
  \begin{equation}
    \label{72.8c}
    |\psi(z,\bar z)|\le c''_1|f(z)|\prod_{j=1}^{n}|\tilde g_j^A(z,\bar z)|
  \end{equation}
  holds true in the interior of $Q$. Formulas (\ref{72.8}),
  (\ref{72.8b}) and (\ref{72.8c}) make it possible to complete the
  proof by arguing in the same way as in the proof of
  Theorem~\ref{orig:thm-6.3}.
\end{proof}

In the case when all chains are parallel we have a somewhat weaker
result.

\begin{thm}\label{orig:thm-7.4}
  Let $\Omega_1,\ldots,\Omega_n$ be parallel chains whose union is a
  uniformly discrete set. Assume that the convergence exponent of
  $\Omega'$ satisfies either $\tau'<1/2$ or $\tau'\le1/2$ and
  $n'(r)=o(r^{1/2})$. Then the Hamiltonians $H_{\max}^{\pm}(\ba')$
  have infinitely degenerate zero modes.
\end{thm}

\begin{rem}
  Under the assumptions of Theorem~\ref{orig:thm-7.4} it holds true
  that $p'=0$ but this equality does not imply the assumptions of the
  theorem.
\end{rem}

\begin{proof}
  In virtue of Lemma~\ref{orig:lem-9.1} one can assume that every
  chain $\Omega_j=K_j+\Lambda_j$, $\Lambda_j=\{k\omega_j;\,k\in\Z\}$,
  is contained in a line $L_j$, with different chains being contained
  in different lines, and that all lines are parallel to the real
  axis.  Hence one can assume that $\omega_j>0$ for all $j$ and that
  all lines $L_j$ are contained in a half-plane $\Im z>a$ where $a>0$.

  Let us consider a function $\psi$ of the form
  \begin{equation}
    \psi(z,\bar z)=f(z)\prod_{j=1}^ng_j(z,\bar z)\prod_{k=1}^m
    |W_k(z)|^{1-\theta'_k}
  \end{equation}
  where
  \begin{equation}
    g_j(z,\bar z)=\prod_{k=1}^{m_j}\left|\sin\!\left(\frac{\pi(z-\kappa_{kj})}
        {\omega_j}\right)\right|^{-\theta_{kj}}
    |V_{kj}(z)|^{\theta_{kj}}\,.
  \end{equation}
  To prove the theorem it suffices to find an infinite, linearly
  independent family of entire functions $f(z)$ such that
  $\psi\in{}L^2(\R^2)$. Let us show that in this case the functions
  \begin{equation}
    \label{72.3}
    f(z)=\frac{\sin(\alpha z)}
    {\sin(\sqrt{\pi\alpha z})\sin(\sqrt{-\pi\alpha z})}\,,
  \end{equation}
  with sufficiently small $\alpha>0$, will do. Here the function
  $f(z)$ is well defined and analytic in the upper half-plane provided
  the usual branch of the square root has been chosen. If $\Im{}z>0$
  then $\sqrt{z}\sqrt{-z}=-iz$. Since
  $\sin(\sqrt{z})/\sqrt{z}=1-z/3!+\ldots$ is in fact an entire
  function, also the function $f(z)$ extends as an entire function.

  We start the verification from several preliminary observations. The
  first one follows from inequality (\ref{6.4a}).

  \medskip

  \noindent{\rm(A)}
  {\it For every $\epsilon>0$ there exists $c>0$ such that}
  \begin{eqnarray*}
    |\sin(\sqrt{z})|\ge \frac{1}{3}\exp(|z|^{1/2})\phantom{-} &&
    \textrm{for } |z|\ge c,\textrm{ }\epsilon\le|\arg z|\le\pi,\\
    |\sin(\sqrt{-z})|\ge \frac{1}{3}\exp(|z|^{1/2}) &&
    \textrm{for } |z|\ge c,\textrm{ }0\le|\arg z|\le\pi-\epsilon.
  \end{eqnarray*}

  \medskip

  Suppose that $n\in\N$ and $0<\delta<\pi/4$. Let us denote
  \begin{displaymath}
    B_n(\delta)=\{z\in\C;\textrm{ }|\sqrt{\pi z}-\pi n|<\delta\}\,.
  \end{displaymath}

  \noindent{\rm(B)} {\it For $n\neq{}m$ the sets $B_n(\delta)$ and
    $B_m(\delta)$ are disjoint}.

  \medskip

  \noindent
  Actually, suppose that $n>m$. If $z\in B_n(\delta)\cap B_m(\delta)$
  then there exist $u,v\in\C$, $|u|,|v|<\delta$, such that $\pi z=(\pi
  n+u)^2=(\pi{}m+v)^2$. Hence $\pi^2(n-m)(n+m)=2\pi(mv-nu)+v^2-u^2$.
  At the same time it holds true that $\pi^2(n-m)(n+m)\ge\pi^2(n+m)$,
  $|2\pi(mv-nu)+v^2-u^2|\le2\pi\delta(m+n)
  +2\delta^2<4\pi\delta(m+n)<\pi^2(n+m)$.

  \medskip

  Set
  \begin{displaymath}
    Q(\epsilon)=\{z\in\C;\,|\sin(\sqrt{\pi z})|<\epsilon\}.
  \end{displaymath}
  Let us denote by $Q_n(\epsilon)$ the connected component of the set
  $Q(\epsilon)$ containing the point $\pi{}n^2$, and by $U(\epsilon)$
  the connected component of the set $\{w\in\C;\textrm{
  }|\sin(w)|<\epsilon\}$ containing $0$. Observe that {\rm(B)} implies
  the following claim.

  \medskip

  \noindent{\rm(C)} {\it For sufficiently small $\epsilon>0$ the
    connected components $Q_n(\epsilon)$ are mutually disjoint}.

  \medskip

  To complete the proof we shall need the following two lemmas.
  \renewcommand{\qed}{}
\end{proof}

\begin{lem}\label{orig:lem-7.2}
  For every $\delta>0$ there exists a constant $c>0$ such that
  $$
  \left|\frac{\sin(z)}{\sin(\sqrt{\pi z})}\right| \le
  c\max(|\sqrt{\pi z}|+1,|\sin(z)|)
  $$
  on the strip $|\Im z|\le\delta$.
\end{lem}

\begin{proof}[Proof of Lemma~\ref{orig:lem-7.2}]
  The equality $|\sin(z)|^2=\sin^2(x)+\sh^2(y)$, with $z=x+iy$,
  $x,y\in\R$, implies that
  $|\sin\,z|\le(x^2+c_\delta^2\,y^2)^{1/2}\le{}c_\delta|z|$ for
  $|\Im{}z|\le\delta$ where we have set $c_\delta=\sh(\delta)/\delta$.

  Let us choose $\epsilon>0$ small enough so that the sets
  $Q_n(\epsilon)$ are mutually disjoint, $|\sqrt{\pi{}z}|>\pi{}n-1$
  for $z\in Q_n(\epsilon)$ and $|\sin(w)|\ge|w|/2$ on $U(\epsilon)$.
  For $z\in\C\setminus Q(\epsilon)$, the desired inequality is valid
  with $c=\epsilon^{-1}$. If $z\in Q_n(\epsilon)$ then
  $\sqrt{\pi{}z}-\pi{}n\in{}U(\epsilon)$ and therefore
  $$
  |\sin(\sqrt{\pi z})|=|\sin\,(\sqrt{\pi z}-\pi n)|
  \ge |\sqrt{\pi z}-\pi n|/2\,.
  $$
  Furthermore, if $|\Im z|\le\delta$ then
  $$
  |\sin(z)|=|\sin(z-\pi{}n^2)|\le c_\delta|z-\pi n^2|.
  $$
  Consequently we have, for $z\in Q_n(\epsilon)$ and $|\Im
  z|\le\delta$,
  $$
  \left|\frac{\sin(z)}{\sin(\sqrt{\pi z})}\right|
  \le 2c_\delta\,\frac{|z-\pi n^2|}{|\sqrt{\pi z}-\pi n|}
  \le \frac{2c_\delta}{\pi}(|\sqrt{\pi z}|+\pi n)
  \le \frac{4c_\delta}{\pi}(|\sqrt{\pi z}|+1)\,.
  $$
  This proves the lemma.
\end{proof}

\begin{lem}\label{orig:lem-7.3}
  For any $b>0$ there exists $\epsilon>0$ such that
  $$
  \iint\limits_{Q(\epsilon)}\left|
    \frac{\sin(z)}{\sin(\sqrt{\pi z})}\right|^2
  \exp\!\big(-b|z|^{1/2}\big)\,dxdy\,<\,\infty
  $$
  where one chooses the principal branch of the square root on
  $\C\setminus\R_-$.
\end{lem}

\begin{proof}[Proof of Lemma~\ref{orig:lem-7.3}]
  We choose $\epsilon$ small enough so that Claim {\rm(C)} is true. In
  the integral
  \begin{displaymath}
    I_n = \iint\limits_{Q_n(\epsilon)}\left|
      \frac{\sin(z)}{\sin(\sqrt{\pi z})}\right|^2
    \exp\!\big(-b|z|^{1/2}\big)\,dxdy
  \end{displaymath}
  we apply the substitution $w=\sqrt{\pi z}-\pi n$, i.e., $z=(w+\pi
  n)^2/\pi$, where $w=u+iv$, $u,v\in\R$. Since
  $dx\wedge{}dy=(4/\pi^2)|w+\pi{}n|^2{}du\wedge{}dv$ and
  $Q_n(\epsilon)$ is mapped onto $U(\epsilon)$ we get
  \begin{displaymath}
    I_n = \frac{4}{\pi^2}\iint\limits_{U(\epsilon)} \left|
      \frac{\sin\!\left(2nw+\frac{w^2}{\pi}\right)}
      {\sin(w)}\right|^2\,|w+\pi n|^2
    \exp\!\left(-\frac{b}{\sqrt{\pi}}|w+\pi n|\right)
    dudv\,.
  \end{displaymath}
  One can assume that $\epsilon$ is sufficiently small so that
  $|\sin(w)|\ge|w|/2$ on $U(\epsilon)$. Furthermore, there clearly
  exists a constant $c>0$, depending on $\epsilon$ but independent of
  $n$, such that $|\sin(2nw+(w^2/\pi))|<c\sh(2n|w|)$ on $U(\epsilon)$.
  Thus we get
  \begin{displaymath}
    I_n\le\frac{16c^2}{\pi^2}\iint\limits_{U(\epsilon)}
    \frac{\sh^2(2n|w|)}{|w|^2}\,|w+\pi n|^2
    \exp\!\left(-\frac{b}{\sqrt{\pi}}|w+\pi n|\right)
    dudv\,.
  \end{displaymath}
  By modifying the constant in front of the integral we can simplify
  this inequality,
  \begin{displaymath}
    I_n \le c'n^2\exp\!\big(-b\,\sqrt{\pi}\,n\big)
    \iint\limits_{U(\epsilon)}
    \frac{\sh^2(2n|w|)}{|w|^2}\, dudv\,.
  \end{displaymath}
  Here again the constant $c'>0$ does not depend on $n$. If
  $\sup\{|w|;\textrm{ }w\in U(\epsilon)\}<b\,\sqrt{\pi}/4$ then
  \begin{displaymath}
    \sum_{n=0}^\infty I_n
    \le c'\iint\limits_{U(\epsilon)}\sum_{n=0}^\infty
    n^2\exp\!\big(-b\,\sqrt{\pi}\,n\big)
    \frac{\sh^2(2n|w|)}{|w|^2}\,dudv < \infty\,.
  \end{displaymath}
  This proves the lemma.
\end{proof}

\begin{proof}[Proof of Theorem~\ref{orig:thm-7.4} (continued)]
  Let us return to the proof of Theorem~\ref{orig:thm-7.4}. We denote
  $A_1=\{z\in\C;\,\Re z\ge 0\}$, $A_2=\{z\in\C;\,\Re z\le 0\}$, and we
  shall show that $\psi\in L^2(A_j)$, $j=1,2$. More precisely, we
  shall prove only the membership $\psi\in L^2(A_1)$, the property
  $\psi\in L^2(A_2)$ can be shown analogously. We split the set $A_1$
  into a union $A_1=P_1\cup P_2$ with
  $P_1=\{z\in{}A_1;\textrm{ }0\le\Im z\le{}b\}$,
  $P_2=A_1\setminus{}P_1$. The bound $b$ is chosen so that the strip
  $P_1$ contains the set $\Omega$ in its interior.

  From the assumptions it follows that the functions $V_{kj}(z)$ and
  $W_j(z)$ are of growth $(1/2,0)$, i.e., for every $\epsilon>0$ there
  exists a constant $C_\epsilon>0$ such that \linebreak
  $\max_{j,k}\{|V_{kj}(z)|,\,|W_k(z)|\}
  \le C_\epsilon\exp(\epsilon|z|^{1/2})$
  (cf. Appendix~\ref{sec:append-analyt-fce}). Let us denote
  \begin{equation}
    \label{73.1}
    G(z,\bar z) = \frac{1}{\sin(\sqrt{-\pi\alpha z})}
    \prod_{l=1}^m|W_l(z)|^{1-\theta'_l}
    \prod_{j=1}^{n}\prod_{k=1}^{m_j}|V_{kj}(z)|^{\theta_{kj}}.
  \end{equation}
  Hence
  \begin{displaymath}
    \psi(z,\bar z) = \frac{\sin(\alpha z)}{\sin(\sqrt{\pi\alpha z})}\,
    G(z,\bar z)\prod\limits_{j=1}^n\tilde g_j(z,\bar z)
  \end{displaymath}
  where
  \begin{displaymath}
    \tilde g_j(z,\bar z) =
    \prod_{k=1}^{m_j}\left|\sin\!\left(
        \frac{\pi(z-\kappa_{kj})}{\omega_j}
      \right)\right|^{-\theta_{kj}}\,.
  \end{displaymath}
  According to Claim {\rm(A)} there exist constants $c_1,c_2>0$ such
  that
  \begin{equation}
    \label{73.2}
    |G(z,\bar z)|\le c_1\exp(-c_2|z|^{1/2})\,,
  \end{equation}
  for all $z\in A_1$.

  With the aid of Lemma~\ref{orig:lem-7.2} and formula (\ref{73.2}) we
  can estimate the function $\psi$ on the strip $P_1$,
  $$
  |\psi(z,\bar z)|\le c'(|\sqrt{\pi\alpha z}|+1)\exp(-c_2|z|^{1/2})
  \prod\limits_{j=1}^n\tilde g_j(z,\bar z)\,.
  $$
  The singularities of $\psi$ in $P_1$ are square integrable and
  therefore, similarly as in the proof of Theorem~\ref{orig:thm-6.3},
  we obtain

  \medskip

  \noindent(1) $\psi\in L^2(P_1)$.

  \medskip

  Since $\Omega\subset{}P_1$, on $P_2\cap\alpha^{-1}Q(\epsilon)$ we
  have the estimate
  $$
  |\psi(z,\bar z)|\le c''
  \left|\frac{\sin(\alpha z)}{\sin(\sqrt{\pi\alpha z})}\right|
    \exp(-c_2|z|^{1/2})
  $$
  If $\epsilon$ is small enough then Lemma~\ref{orig:lem-7.3} implies that

  \medskip

  \noindent(2) $\psi\in L^2(P_2\cap\alpha^{-1}Q(\epsilon))$.

  \medskip

  Finally, for $z\in P_2\setminus\alpha^{-1}Q(\epsilon)$ we have
  $|\sin(\sqrt{\pi\alpha z})|\ge\epsilon$ and hence the inequality
  $$
  |\psi(z,\bar z)|\le c''' |\sin(\alpha z)|
  \prod\limits_{j=1}^n\tilde g_j(z,\bar z)
  $$
  holds true. We conclude, similarly as in the proof of
  Theorem~\ref{orig:thm-6.3}, that if
  \begin{displaymath}
    0 < \alpha < \sum_{j=1}^n\sum_{k=1}^{m_j}\frac{\pi}{\omega_j}
    \theta_{kj}
  \end{displaymath}
  then

  \medskip

  \noindent(3) $\psi\in L^2(P_2\setminus Q_\alpha(\epsilon))$.

  \medskip

  \noindent This concludes the proof of the theorem.
\end{proof}

\subsection{Addition of solenoids to an Aharonov--Bohm lattice}

The case when the Aharonov--Bohm fluxes are arranged in a lattice
$\Omega$ has been discussed in Theorem~\ref{orig:thm-6.5}. It turns
out that that for a scarce perturbation the result stated in the
theorem is still true.

\begin{thm}
  Suppose that $\Omega$ is a lattice, i.e.,
  $\Omega_j=\kappa_j+\Lambda$ where $\Lambda$ is a Bravais lattice of
  rank 2. Suppose further that the genus of the set $\Omega'$ fulfills
  $p'\le1$. Then the Hamiltonians $H_{\max}^{\pm}(\ba')$ have
  infinitely degenerate zero modes.
\end{thm}

\begin{proof}
  Let $\tilde{W}(z)$ be the Weierstrass canonical product for the set
  $\Omega'$, let $\varrho'$ be the growth order of $\tilde{W}(z)$, and
  let $\tau'$ be the convergence exponent of $\Omega'$. Then
  $\varrho'=\tau'\leq{}p'+1\leq2$ (see
  Appendix~\ref{sec:append-analyt-fce}). If $\varrho'<2$ then
  $\tilde{W}(z)$ is of growth $(2,0)$. If $\varrho'=2=p'+1$ then, by
  Theorem~\ref{orig:thm-9.5} (b), $\tilde{W}(z)$ is of minimal type.
  Consequently, also in the latter case $\tilde{W}(z)$ is of growth
  $(2,0)$.  This means that for any $c>0$ there exist $a>0$ and $R>0$
  such that for all $z$, $|z|>R$, it holds true that
  \begin{equation}
    \label{eq:growth-W'(z)}
    |\tilde{W}(z)|\leq a\,\exp\big(c|z|^2\big).
  \end{equation}
  The same observation is clearly true for any subset of $\Omega'$. In
  particular, the functions $V_j(z)$ and $W_\ell(z)$ are of growth
  $(2,0)$ and obey estimates similar to (\ref{eq:growth-W'(z)}).

  Zero modes of $H_{\max}^{+}(\ba')$ are gauge equivalent to functions
  of the form
  \begin{displaymath}
    \psi(z,\bar{z}) = f(z)
    \prod_{j=1}^n|\tilde\sigma(z-\kappa_j)|^{-\theta_j}
    \prod_{j=1}^n|V_j(z)|^{\theta_j}
    \prod_{j=1}^m|W_\ell(z)|^{1-\theta'_\ell}\,.
  \end{displaymath}
  From Lemma~\ref{orig:lem-6.3} and from the estimate
  (\ref{eq:growth-W'(z)}) with sufficiently small $c>0$ it follows
  that $\psi$ is square integrable if $f(z)$ is an arbitrary
  polynomial.
\end{proof}

\subsection{Irregular systems of Aharonov--Bohm solenoids}

All the preceding results were concerned with Aharonov--Bohm systems
$\Omega$ with bounded density, i.e., for which
$\limsup_{r\to\infty}\,n(r)/r^2<\infty$ (cf. (~\ref{6.7.3})).  Here we
show that zero modes may occur also in systems with infinite density,
more precisely, in systems for which
$\liminf\limits_{r\to\infty}\,n(r)/r^2=\infty$.  Moreover, we shall
present examples of systems $\Omega$ with arbitrarily large
convergence exponent $\tau_\Omega$. Let us fix a natural number
$N\ge2$ and set
$$
\Omega_N=\{e^{\pi i k/N}\,m^{1/N};\textrm{ }m\in\N\,,
k=0,1,\ldots,2N-1\}\,.
$$
Obviously, the convergence exponent $\tau$ for the set $\Omega_N$
equals $N$. In particular, for $N>2$ we have
$\lim_{r\to\infty}\,n(r)/r^2=\infty$. Let $\theta$ be an arbitrary
number from the interval $]0,1[$. Then the vector potential $\ba$ of
the Aharonov--Bohm system determined by the couple $(\Omega_N,\theta)$
reads $\ba(z,\bar{z})=\theta\,\sgrad\,\ln(|W(z)|)$ where
$$
W(z)=\frac{\sin(\pi z^N)}{z^{N-1}}\,.
$$

\begin{thm}
  The Hamiltonians $H_{\max}^{\pm}(\ba)$ have infinitely degenerate
  zero modes.
\end{thm}

\begin{proof}
  It is sufficient to show that for $0<\alpha<\pi\theta$ the function
  $$
  \psi(z,\bar z)=\frac{\sin(\alpha z^N)}{z^N}\,|W(z)|^{-\theta}
  $$
  is square integrable. Set
  $$
  S=\left\{z\in\C;\textrm{ }-\frac{\pi}{2N}
    < \arg z<\frac{3\pi}{2N}\right\}\,.
  $$
  Then
  $$
  \iint\limits_{\R^2}|\psi(z,\bar z)|^2\,dxdy
  = N\iint\limits_{S}|\psi(z,\bar z)|^2\,dxdy
  $$
  and therefore it suffices to verify that
  \begin{equation}
    \label{76.1}
    \iint\limits_{S}|\psi(z,\bar z)|^2\,dxdy\,<\,\infty\,.
  \end{equation}
  Let us make a substitution of the integration variable in
  (\ref{76.1}), $w=z^N$ where $w=u+iv$, $u,v\in\R$. Since
  $du\wedge dv=N^2|z|^{2N-2}dx\wedge dy$ we can rewrite the integral as
  \begin{equation}
    \label{76.2}
    \iint\limits_{S}|\psi(z,\bar z)|^2\,dxdy\,=
    \frac{1}{N^2}\iint\limits_{\R^2}\frac{1}{|w|^\beta}
    |\sin(\alpha w)|^2|\sin(\pi w)|^{-2\theta}\,dudv
  \end{equation}
  where
  $$
  \beta=4-2\theta-2\frac{1-\theta}{N}\,.
  $$
  Since $2-2\theta-\beta>-2$ and $\beta>1$ one can show that
  the integral (\ref{76.2}) is finite using a reasoning as in the
  proof of Theorem~\ref{orig:thm-6.3}.
\end{proof}


\appendix

\makeatletter

\newcommand\appsection{\@startsection {section}{1}{\z@}%
{-3.5ex \@plus -1ex \@minus -.2ex}%
{2.3ex \@plus.2ex}%
{\normalfont\Large\bfseries\noindent%
Appendix~\thesection.\hspace{0.5em}}}

\renewcommand{\@seccntformat}[1]{}

\makeatother

\appsection{Lattices}

Here we collect basic definitions and some auxiliary results about
lattices. Let $E$ be a finite-dimensional real Euclidean space with
dimension $d$. A discrete subgroup $\Lambda$ of the additive group $E$
is called {\it a Bravais lattice}. For any Bravais lattice $\Lambda$
there exist linearly independent vectors
$\omega_1\,,\ldots,\omega_r\in E$ such that
$$
\Lambda=\Z\omega_1+\Z\omega_2+\ldots+\Z\omega_r\,.
$$
The array $(\omega_j)_{1\le j\le r}$ is called {\it a basis} of the
Bravais lattice $\Lambda$. The integer $r$ does not depend on the
choice of basis and is called {\it the rank} of the lattice $\Lambda$.
To every basis $(\omega_j)_{1\le j\le r}$ one relates {\em the
  elementary cell} $F$, $F\subset E$, formed by all $x\in E$ whose
orthogonal projection $x'$ onto the linear span $L$ of the lattice
$\Lambda$ has a decomposition
$$
x'=t_1\omega_1+t_1\omega_2\ldots+t_r\omega_r\,,
$$
with $0\le t_j<1$ for all $j$. If $r=d$ (the dimension of the
lattice $\Lambda$ is maximal possible) then $F$ is a convex
parallelepiped. In the opposite case $F$ is even not bounded.

A non-empty discrete subset $\Omega\subset E$ is called {\it a
  crystal} with the Bravais lattice $\Lambda$ if it is invariant with
respect to the action of $\Lambda$ on $E$ and has a finite number
of orbits. Obviously, every crystal $\Omega$ can be written in the
form $\Omega=K+\Lambda$ where $K\subset E$ is a finite set whose
number of elements equals the number of orbits. Without loss of
generality we may assume that $K\subset F$. Conversely, every set
of the form $\Omega=K+\Lambda$ is a crystal. If $|K|=1$ then the
crystal is called {\it mono-atomic} or {\it simple}. In the
general case when $|K|=n$ the crystal $\Omega$ is called {\it
$n$-atomic}.  If $r=1$ then $\Omega$ is called {\it a chain} ({\it
a simple chain} if in addition $|K|=1$). If $r=d$ then $\Omega$ is
called {\it a lattice} (more precisely, {\it
  a crystal lattice}) in the space $E$.  In other words, a crystal is
such a discrete subset $\Omega\subset E$ whose group of parallel
translations acts co-compactly on $E$.

Let us note that in our definition we do not exclude the case $r=0$.
If so then $\Lambda=\{0\}$ and a crystal with the Bravais lattice
$\Lambda$ is simply a finite subset of $E$.

It is worth of noticing that a crystal $\Omega$ is always {\it a
  uniformly discrete} subset of $E$. This means that there exists a
constant $c>0$ such that $|\omega'-\omega''|\ge c$ whenever
$\omega',\omega''\in\Omega$, $\omega'\ne\omega''$.

We shall need the following lemma.

\begin{lem}\label{orig:lem-9.1}
  Assume that $\dim E=1$ and that $\Omega_1\,,\ldots, \Omega_n$
  are chains in $E$. The union $\Omega=\Omega_1\cup\ldots\cup\Omega_n$
  is a chain if and only if $\Omega$ is a uniformly discrete set.
\end{lem}

\begin{proof}
  We only need to prove that this condition is sufficient. Moreover,
  it suffices to consider the case $n=2$. The general case then
  follows by mathematical induction. Let us write
  $$
  \Omega_j=K_j+\Lambda_j\quad (j=1,2)\,,
  $$
  where $\Lambda_j$ is the Bravais lattice of $\Omega_j$. Let us
  identify $E$ with $\R$. Then $\Lambda_j=\Z\omega_j$, with
  $\omega_j>0$.

  We shall show that the number $p=\omega_1/\omega_2$ is rational.
  Actually, in the opposite case the set $\Z\omega_1+\Z\omega_2$ would
  be dense in $\R$. Let us choose $\kappa_1\in K_1$ and
  $\kappa_2\in K_2$. We can find a sequence
  $n_k\omega_2-m_k\omega_1$ ($m_k\,,n_k\in\Z$) converging to
  $\kappa_1-\kappa_2$ and such that $\kappa_1-\kappa_2\ne
  n_k\omega_2-m_k\omega_1$ for all $k$. Obviously, this contradicts
  the assumption that the set $\Omega$ is uniformly discrete.

  Hence $p=N/M$, with $N,M\in\N$. Then $M\omega_1=N\omega_2$ and
  therefore $\Omega$ is invariant with respect to the lattice
  $\Lambda$ with the basis vector $M\omega_1$. It is easy to see that
  the number of orbits of the group $\Lambda$ in $\Omega$ is finite.
  But this means that $\Omega$ is a chain.
\end{proof}

\begin{rem}\label{orig:rem-9.1}
  For $\dim E>1$ an analog of Lemma~\ref{orig:lem-9.1} is false.
  Actually, already for $\dim E=2$ it can happen that a union of two
  simple lattices is uniformly discrete but not a lattice. This is
  demonstrated by the following example. Let $E=\R^2$. Let $\Lambda_1$
  be a Bravais lattice with the basis $\omega_1={\bf e}_1$,
  $\omega_2={\bf e}_2$, and let $\Lambda_2$ be a Bravais lattice with
  the basis $\omega'_1=\sqrt{2}{\bf e}_1$, $\omega'_2={\bf e}_2$.
  Consider the crystal lattices $\Omega_1=\Lambda_1$ and
  $\Omega_2=\kappa+\Lambda_2$ where $\kappa=(1/2){\bf{e}}_2$.
  Obviously, $\Omega=\Omega_1\cup\Omega_2$ is a uniformly discrete
  set. Let $\Lambda$ be a group of parallel translations acting on
  $\Omega$.  Suppose that the number of orbits of the group $\Lambda$
  in $\Omega$ is finite. Then there exist $n\,,m\in\Z$, $n\ne m$, such
  that $n\omega_1$ and $m\omega_1$ belong to the same orbit. Hence
  $k\omega_1\in\Lambda$ for some $k\in\Z$, $k\ne 0$. But
  $\kappa+k\omega_1\notin\Omega$.
\end{rem}

\appsection{Auxiliary results from the theory of analytic functions}
\label{sec:append-analyt-fce}

Here we recall some results from the theory of analytic functions that
are necessary for our presentation, for the details see
\cite{Mar,Mar2,Boa}. For an entire function $f$ we set
\begin{equation}
  \label{8.2}
  M_f(r)\equiv
  M(r)=\max\limits_{|z|=r}\,|f(z)|=\max\limits_{|z|\le r}\,|f(z)|\,.
\end{equation}
{\it The order} (more precisely, {\it the growth order}) of an entire
function $f$ is the number
\begin{equation}
  \label{8.3}
  \varrho_f\equiv\varrho
  = \inf\{\alpha;\textrm{ }\exists R_\alpha>0\textrm{ }
  \forall r>R_\alpha \textrm{ } M(r)<\exp(r^\alpha)\}\,,
\end{equation}
or, equivalently,
\begin{equation}
  \label{8.4}
  \varrho_f =
  \limsup\limits_{r\to\infty}\,\,\,
  \frac{\ln\!\big(\ln(M(r))\big)}{\ln(r)}\,.
\end{equation}
Let us note that if $f(z)=\sum_{n=0}^{\infty}a_nz^n$ then
\begin{equation}
  \label{8.5}
  \varrho_f =
  \limsup\limits_{n\to\infty}\,\,\,
  \frac{\ln(n)}{\ln\,(|a_n|^{-1/n})}\,.
\end{equation}

If $\varrho_f<\infty$ then one says that $f$ is a function {\it of
  finite order}. For a function of finite order $\varrho$ the number
\begin{equation}
  \label{8.6}
  \varsigma_f\equiv\varsigma=
  \inf\{K>0;\textrm{ }\exists R_K>0\textrm{ }\forall r>R_K
  \textrm{ }M(r)<\exp(Kr^\varrho)\}
\end{equation}
is well defined and it is called {\it the type of the function} $f$.
The type can be equivalently defined by the formula
\begin{equation}
  \label{8.7}
  \varsigma_f = \limsup\limits_{r\to\infty}\,\,
  \frac{\ln\!\big(M(r)\big)}{r^\varrho}\,.
\end{equation}
Moreover, $\varsigma_f$ can be expressed in terms of the Taylor
coefficients $a_n$,
\begin{equation}
  \label{8.8}
  (\varsigma_f e\varrho_f)^{1/\varrho}=\limsup\limits_{n\to\infty}\,\,
  n^{1/\varrho}\,|a_n|^{1/n}\,.
\end{equation}
If $\varrho_f<\infty$ and $\varsigma_f=0$ then $f$ is called a
function of {\it minimal type}.

A function $f$ of finite order $\varrho$ and of finite type
$\varsigma$ obeys the estimate
\begin{equation}
  \label{8.9}
  |f(z)|\le\exp((\varsigma+\epsilon)|z|^\varrho)
\end{equation}
for arbitrary $\epsilon>0$ and $|z|$ greater than a constant
$R_\epsilon>0$.
Conversely, if the estimate
\begin{equation}
  \label{8.10}
  |f(z)|\le c\,\exp(\varsigma_1|z|^{\varrho_1})
\end{equation}
is fulfilled with a constant $c>0$ then the function $f$ has both a
finite order and a finite type, and it holds $\varrho_f\le\varrho_1$
and $\varsigma_f\le\varsigma_1$. A couple of numbers
$(\varrho_1,\varsigma_1)$, $0\le\varrho_1,\varsigma_1\le\infty$,
determines {\it the growth} of a function $f(z)$ if
$\varrho_f\le\varrho_1$ and $\varrho_f=\varrho_1$ implies
$\varsigma_f\le\varsigma_1$. Functions of growth $(1,\varsigma_1)$,
with $\varsigma_1<\infty$, are said to have {\it exponential growth}.

The order and the type of an entire function on one side and the
distribution of its zeroes on the other side are deeply related. Let
us arrange all nonzero elements of a discrete set $\Omega\subset\C$ in
a sequence $\Omega_*=(\omega_k)_{k\ge1}$ which is ascending in the
absolute value and ascending in the argument ($0\le\arg\,z<2\pi$) in
the case of equal absolute values. {\it The convergence exponent} of
the set $\Omega$ (or of the sequence $\Omega_*$) is the number
\begin{equation}
  \label{3.5}
  \tau_\Omega\equiv
  \tau=\inf\left\{\alpha>0;\textrm{ }\sum\limits_{k=1}^{\infty}
    \frac{1}{|\omega_k|^\alpha}\,<\,\infty\right\}\,,
\end{equation}
or, equivalently,
\begin{equation}
  \label{3.6}
  \tau_\Omega = \limsup\limits_{k\to\infty}\,\,\,
  \frac{\ln(k)}{\ln(|\omega_k|)}\,.
\end{equation}
If $\tau_\Omega$ is finite then the number
\begin{equation}
  \label{3.7}
  p_\Omega\equiv p=
  \begin{cases}
    \max\left\{n\in\N;\textrm{ }\sum\limits_{k=1}^{\infty}\frac{1}
      {|\omega_k|^n}=\infty\right\} &
    \textrm{if } \Omega \textrm{ is infinite}, \cr
    \noalign{\medskip}
    -\infty & \textrm{if } \Omega \textrm{ is finite}, \cr
  \end{cases}
\end{equation}
is well defined and it is called {\it the genus} of the set $\Omega$
(or of the sequence $\Omega_*$). For $\tau_\Omega=\infty$ we set
$p_\Omega=\infty$.

For $r>0$ we set
\begin{equation}
  \label{8.11}
  n_{\Omega}(r)\equiv n(r)=\#\{\omega\in\Omega;\textrm{ }|\omega|\le r\}\,.
\end{equation}
For a non-empty set $\Omega$ the formula
\begin{equation}
  \label{2A.2}
  \tau_\Omega = \limsup\limits_{r\to\infty}\,
  \frac{\ln\!\big(n(r)\big)}{\ln(r)}\,,
\end{equation}
shows that that the convergence exponent of an discrete set
characterizes its density \cite[Theorem 2.5.8]{Boa}.

On the other hand, let $\Omega_f\equiv\Omega$ be the zero set of an
entire function $f$. Then the following fundamental inequality is
valid (the Hadamard theorem, see for example \cite[Theorem
2.5.18]{Boa}):
\begin{equation}
  \label{2A.2a}
  \tau_\Omega\le \varrho_f\,.
\end{equation}
If $\varrho_f$ is not an integer then $\tau_\Omega=\varrho_f$
\cite[Theorem 2.9.1]{Boa}. From (\ref{2A.2a}) we deduce that
\begin{equation}
  \label{2A.3}
  n(r)=O(r^{\varrho+\epsilon})
\end{equation}
for any $\epsilon>0$. If $\varrho_f>0$ and $\varsigma_f<\infty$ then a
stronger estimate is valid \cite[Theorem 2.5.13]{Boa}:
$$
L\equiv \limsup\limits_{r\to\infty}\,r^{-\varrho}n(r)\le e\varrho_f\,
\varsigma_f\,,
$$
\begin{equation}
  \label{2A.4}
  l\equiv \liminf\limits_{r\to\infty}\,r^{-\varrho}n(r)\le \varrho_f\,
\varsigma_f\,.
\end{equation}
Moreover, if $\tau_\Omega>0$ then $Le^{l/L}\le\varrho_f\varsigma_f$,
in particular, $L+l\le\varrho_f\varsigma_f$.

For $\varrho_f<\infty$ it can never happen that
$n_\Omega(r)=o(r^{\varrho-\epsilon})$ \cite[Theorem 2.9.3]{Boa}. The
following theorem is due to Lindel\"of (see \cite[Theorem 2.9.5 and
Theorem 2.10.1]{Boa}).

\begin{thm}\label{orig:thm-9.1}
  (1) Assume that $\varrho\equiv\varrho_f<\infty$ is not an integer.
  An entire function $f(z)$ is of finite type if and only if
  $n_\Omega(r)=O(r^{\varrho})$, and it is of minimal type if and only
  if $n_\Omega(r)=o(r^{\varrho})$.

  \noindent(2) Assume that $\varrho_f$ is a positive integer. The
  function $f(z)$ is of finite type if and only if
  $n_\Omega(r)=O(r^{\varrho})$ and the sums
  $$
  S(r)=\sum\limits_{0<|\omega_k|\le r} \omega_k^{-\varrho}
  $$
  are bounded.
\end{thm}

An entire function with simple zeroes is determined by its zero
set up to an multiplier $e^{g(z)}$ where $g(z)$ is an entire
function. Furthermore, for an arbitrary discrete set
$\Omega\subset\C$ there exists an entire function $f(z)$ with
simple zeroes whose zero set coincides with $\Omega$. Denote by
$E(u,p)$ the Weierstrass canonical multiplier, with $u\in\C$ and
with $p\in\N$,
$$
E(u,p)=(1-u)\exp\!\left(u+\frac{u^2}{2}+\ldots+\frac{u^p}{p}\right)
$$
(by definition, $E(u,0)=1-u$). Let $\Omega_*=(\omega_k)_{k\ge 1}$
be, as above, the sequence formed by all nonzero elements of the set
$\Omega$ appropriately enumerated. Let us denote by $\chi_\Omega\equiv
\chi$ an integer which is equal 1 if $0\in\Omega$ and 0 in the
opposite case. {\it The Weierstrass canonical product} associated to
$\Omega$ is by definition an entire function $W_\Omega(z)$ defined by
the infinite product
\begin{equation}
  \label{2A.5}
  z^\chi\prod\limits_{k=1}^{\infty}E(z/\omega_k,p_\Omega)
\end{equation}
if the convergence exponent of $\Omega$ is finite, and by the infinite
product
\begin{equation}
  \label{2A.6}
  z^\chi\prod\limits_{k=1}^{\infty}E(z/\omega_k,k)
\end{equation}
in the opposite case.

\begin{thm}[Weierstrass, Hadamard]
  The infinite product defining the function $W_\Omega(z)$ converges
  absolutely and locally uniformly. Consequently, $W_\Omega(z)$ is an
  entire function and its zero set coincides with $\Omega$. Moreover,
  the zero set of an entire function $f(z)$ with simple zeroes only
  equals $\Omega$ if and only if the function $f(z)$ is of the form
  \begin{equation}
    \label{9.1a}
    f(z)=e^{g(z)}\,W_\Omega(z)
  \end{equation}
  where $g(z)$ is an entire function. The growth order of the function $W_\Omega(z)$ equals the
convergence exponent of the set $\Omega$.
\end{thm}

Moreover, the following theorem is true.

\begin{thm}[Hadamard]
  If the function $f$ from relation (\ref{9.1a}) has a finite order
  $\varrho_f$ then $g(z)$ is a polynomial with a degree not exceeding
  $[\varrho_f]$.
\end{thm}

\begin{thm}[Borel Theorem]\label{orig:thm-9.4}
  Conversely, if $\tau_\Omega<\infty$ and $f(z)$ is a function written
  in the form (\ref{9.1a}) where $g(z)$ is a polynomial of degree $n$
  then $f(z)$ has a finite order $\varrho_f=\max(\tau,n)$. If either
  $\tau_\Omega<n$ or the series
  $\sum_{k=1}^{\infty}|\omega_k|^{-\tau}$ is convergent then the
  function $f(z)$ is of finite type.
\end{thm}

{\it The genus} of a function $f(z)$ having the form (\ref{9.1a}),
where $g(z)$ is a polynomial of degree $n$, is the integer
$q_f\equiv{}q=\max(n,p_\Omega)$. The following theorem is a useful
completion of Theorem \ref{orig:thm-9.1} due to Lindel\"of
\cite[Theorem 2.10.3]{Boa}.

\begin{thm}\label{orig:thm-9.5}
  Under the assumptions of Theorem \ref{orig:thm-9.1} let $\varrho_f$
  be a positive integer. A function $f(z)$ written in the form
  (\ref{9.1a}), where $g(z)$ is a polynomial, is of minimal type if
  and only if one of the following conditions is satisfied:

  \noindent$($a$)$ $n_\Omega(r)=o(r^{\varrho})$, $p_\Omega=\varrho_f$,
  and
  $$
  \sum\limits_{k=1}^{\infty} \omega_k^{-\varrho}=-\varrho_f\alpha_0
  $$
  where $\alpha_0$ is the coefficient (possibly vanishing) standing
  at $z^{\varrho}$ in the polynomial $g(z)$,

  \noindent$($b$)$ $p_\Omega=\varrho_f-1$ and $\alpha_0=0$.

  In particular, if $q_f<\varrho_f$ then $f(z)$ is of minimal type.
\end{thm}

We shall also need the following particular case of the Mittag-Leffler
theorem (see \cite[II.7.3.2]{Mar} or \cite{Mar2}).

\begin{thm}
  For an arbitrary discrete subset $\Omega$ of the complex plane $\C$
  and for an arbitrary sequence of complex numbers
  $(\theta_\omega)_{\omega\in\Omega}$ there exists a meromorphic
  function $M(z)$ obeying the following conditions:

  \noindent$(1)$ $M(z)$ has only simple poles,

  \noindent$(2)$ the set of poles of the function $M(z)$ coincides with
  $\Omega$,

  \noindent$(3)$ the residuum of $M(z)$ at the point $\omega$ equals
  $\theta_\omega$.
\end{thm}

\appsection{The Weierstrass $\sigma$-function and related functions}
\label{sec:weierstr-sigma-funct}

In our approach an important role is played by the order $\varrho$ and
by the type $\varsigma$ of the Weierstrass $\sigma$-function
$\sigma(z)$,
$$
\sigma(z)\equiv\sigma(z;\omega_1,\omega_2)
= z\prod\limits_{\omega\in\Lambda\setminus\{0\}}
\left(1-\frac{z}{\omega}\right)
\exp\!\left(\frac{z}{\omega}+\frac{z^2}{2\omega^2}\right)\,.
$$
It is easy to see that the convergence exponent of any lattice in
the plane equals 2. Actually, the series
\begin{equation}
  \label{9.4}
  \mathop{\sum{}^{{}'}}\limits_{n_1,n_2=-\infty}^{+\infty}
  \frac{1}{|n_1\omega_1+n_2\omega_2|^\alpha}
\end{equation}
(the dash indicates, as usual, that the summand with indices
$n_1=n_2=0$ is omitted) converges if and only if $\alpha>2$.
Hence, by the Borel theorem, $\varrho=2$. Since for $\alpha=2$ the
series (\ref{9.4}) diverges the Borel theorem does not say
anything about the type of the function $\sigma(z)$ (apart of the
fact that it is finite). The type of this function has been found
in the general case by A.~M. Perelomov \cite{Per}. In order to
make our presentation self-contained we reproduce below some
details from his derivation.

Let us start from recalling the notation
\begin{equation}
  \label{10.1}
  \eta_j=2\,\zeta\!\left(\frac{\omega_j}{2}\right)
\end{equation}
and the fact that the $\sigma$-function is quasi-periodic in the
following sense:
\begin{equation}
  \label{10.2}
  \sigma(z+\omega_j)=-\sigma(z)
  \exp\!\left(\eta_j\!\left(z+\frac{\omega_j}{2}
    \right)\right)\,.
\end{equation}
Recall also that $S=\Im(\bar\omega_1\omega_2)$ designates the area of
the elementary cell.

\begin{lem}[\cite{Per}]\label{orig:lem-9.2}
  The function $|\sigma(z)|^2$ can be expressed in the form
  \begin{equation}
    \label{10.3}
    |\sigma(z)|^2=\exp(\nu z^2+\bar\nu\bar z^2+2\mu z\bar z)\rho(z,\bar z)
  \end{equation}
  where $\rho$ is a $\Lambda$-periodic function,
  \begin{equation}
    \label{10.4}
    \nu=\frac{i}{4S}(\eta_1\bar\omega_2-\eta_2\bar\omega_1)\,,\textrm{ }
    \mu=\frac{\pi}{2S}\,.
  \end{equation}
\end{lem}

\begin{proof}
  From (\ref{10.2}) we obtain
  \begin{equation}
    \label{10.5}
    |\sigma(z+\omega_j)|^2=|\sigma(z)|^2
    \exp\!\left(2\Re\!\left(\eta_j\left(z+\frac{\omega_j}{2}
        \right)\right)\right)\,.
  \end{equation}
  On the other side, the function $\rho$ defined by equality
  (\ref{10.3}), with $\nu\in\C$ and $\mu\in\R$, is periodic if and
  only if it holds
  \begin{equation}
    \label{10.7}
    |\sigma(z+\omega_j)|^2=
    \exp\!\left(2\nu z\omega_j+2\overline{\nu z\omega_j}+2\mu
      (z\bar\omega_j+\bar z\omega_j)+\nu\omega_j^2+\bar\nu\bar\omega_j^2
      +2\mu\omega_j\bar\omega_j\right)|\sigma(z)|^2\,.
  \end{equation}
  Comparing (\ref{10.5}) to (\ref{10.7}) and taking into account the
  equality
  $$
  2\Re\eta_j\!\left(z+\frac{\omega_j}{2}\right)=
  \eta_jz+\bar\eta_j\bar z+\frac{\eta_j\omega_j}{2}+
  \frac{\overline{\eta_j\omega_j}}{2}\,,
  $$
  we arrive at the system
  \begin{equation}
    \label{10.8}
    \begin{aligned}
      & \nu\omega_j+\mu\bar\omega_j = \frac{1}{2}\eta_j\,,
      & j=1,2\,,\\
      \noalign{\medskip}
      & \nu\omega_j^2+\bar\nu\bar\omega_j^2+2\mu\omega_j\bar\omega_j
      = \frac{1}{2}(\eta_j\omega_j+\bar\eta_j\bar\omega_j)\,,
      & j=1,2\,.
    \end{aligned}
  \end{equation}
  The first couple of equations in (\ref{10.8}) gives
  \begin{equation}
    \label{10.9}
    \nu = \frac{1}{2}\,\frac{\eta_1\bar\omega_2-\eta_2\bar\omega_1}
    {\omega_1\bar\omega_2-\bar\omega_1\omega_2}\,,
    \quad
    \mu = \frac{1}{2}\,\frac{\omega_1\eta_2-\omega_2\eta_1}
    {\omega_1\bar\omega_2-\bar\omega_1\omega_2}\,.
  \end{equation}
  Since $\omega_1\bar\omega_2-\bar\omega_1\omega_2=-2iS$ and in virtue
  of the Lagrange identity
  \begin{equation}
    \label{10.10}
    \eta_1\omega_2-\eta_2\omega_1=2\pi i
  \end{equation}
  we find that relations~(\ref{10.9}) and (\ref{10.4}) are equivalent.
  Using (\ref{10.9}) and the fact that $\mu$ is real one can check
  that the second couple of equations in (\ref{10.8}) is satisfied
  identically.
\end{proof}

\begin{lem}[\cite{Per}]\label{orig:lem-9.3}
  The type of the function $\sigma(z;\omega_1,\omega_2)$ is given by
  the equality
  \begin{equation}
    \label{11.1}
    \varsigma=|\nu|+\mu=\frac{1}{4S}\left(|\eta_1\bar\omega_2-\eta_2\bar
      \omega_1|+2\pi\right)\,.
  \end{equation}
\end{lem}

\begin{proof}
  Let us rewrite (\ref{10.3}) as follows,
  \begin{equation}
    \label{11.2}
    |\sigma(z)|^2 = \exp\!\left((\nu+\bar\nu)(x^2-y^2)+2i(\nu-\bar\nu)xy
      +2\mu(x^2+y^2)\right)\rho(z,\bar z)\,.
  \end{equation}
  The quadratic form occurring in the exponent,
  $2\Re(\nu)(x^2-y^2)-4\Im(\nu)\,xy$, can be diagonalized with the aid
  of a rotation of the coordinate system. Eigenvalues of the
  corresponding symmetric matrix are $\lambda_1=-\lambda_2=2|\nu|$.
  Set $\epsilon=e^{i\phi}$ where $\phi$ is the angle of the rotation.
  Since the quadratic form $x^2+y^2$ is rotationally invariant we have
  \begin{equation}
    \label{11.6a}
    |\sigma(\epsilon z)|=\exp\left((\mu+|\nu|)x^2+(\mu-|\nu|)y^2\right)
    \rho^{1/2}(\epsilon z,\bar\epsilon\bar z)\,.
  \end{equation}
  Owing to the periodicity of the function $\rho$ it holds true that
  \begin{equation}
    \label{11.7}
    \max\limits_{|z|=r}|\sigma(z)|=
    \max\limits_{|z|=r}|\sigma(\epsilon z)|\le
    c\exp\left((\mu+|\nu|)r^2\right)\,.
  \end{equation}
  Consequently, $\varsigma\le |\nu|+\mu$.

  To show the opposite inequality it suffices to construct a sequence
  $z_k$ such that $|z_k|\to\infty$ and
  \begin{equation}
    \label{11.7b}
    |\sigma(\epsilon z_k)|
    \ge c\exp((\mu+|\nu|-\delta_k)|z_k|^2)\,,
  \end{equation}
  where $\delta_k\downarrow 0$ and $c>0$ is a fixed constant. First we
  note that, by the uniqueness theorem for analytic functions in a
  real variable, there exists a point $z_0$ such that
  $\rho(z_0,\bar{z}_0)\ne0$. Then there exists $c>0$ such that
  $|\rho(z,\bar z)|>c$ on a neighborhood $V$ of $z_0$. This gives the
  choice of $c$. Further we consider the canonical mapping
  $h:\,\R^2\longrightarrow\R^2/\Lambda$. Two cases are possible:
  either the image $h(z_0+\R)$ is a closed curve in the torus
  $T=\R^2/\Lambda$ or this image is dense in $T$. In the former case
  there exists a sequence $\lambda_k\in\R$ such that
  $\lambda_k\to\infty$ and
  \begin{equation}
    \label{11.7a}
    h(z_0+\lambda_k)=h(z_0)\,,
  \end{equation}
  in the latter case condition (\ref{11.7a}) should be replaced by
  $h(z_0+\lambda_k)\to h(z_0)$. In the both cases condition
  (\ref{11.7b}) holds true with $z_k=z_0+\lambda_k$.
\end{proof}

Following \cite{Per} we introduce the function
$$
\tilde\sigma(z)=e^{-\nu z^2}\sigma(z).
$$
Lemma~\ref{orig:lem-9.2} implies the equality
\begin{equation}
  \label{0.2}
  |\tilde\sigma(z)|^2=\exp(2\mu|z|^2)\rho(z,\bar z)\,.
\end{equation}

\begin{lem}[\cite{Per}]
  Let $f(z)$ be an entire function whose zero set coincides with
  $\Lambda=\Z\omega_1+\Z\omega_2$, with all zeroes being simple. Then
  the order $\varrho_f$ is at least $2$, $\varrho_f\ge 2$, and if
  $\varrho_f=2$ then the type $\varsigma_f$ is at least $\mu$,
  $\varsigma_f\ge\mu=\pi/2S$. Moreover, in the case of the function
  $\tilde\sigma(z)$ the minimal values are achieved both for the order
  $\varrho$ and the type $\varsigma$, i.e., $\varrho_{\tilde\sigma}=2$
  and $\varsigma_{\tilde\sigma}=\mu=\pi/2S$.
\end{lem}

\begin{proof}
  Since the function $\sigma(z)$ is expressed as a Weierstrass
  canonical product its order equals the convergence exponent
  $\tau_\Lambda=2$.  Let us consider the entire function
  $f(z)=e^{-\alpha z^2}\sigma(z)$, with $\alpha\in\C$. Then
  \begin{equation}
    |f(z)|^2 = \exp\!\Big(2\Re(\nu-\alpha)(x^2-y^2)-4\Im(\nu-\alpha) xy
    +2\mu(x^2+y^2)\Big)\,\rho(z,\bar z)\,.
  \end{equation}
  It is clear that the order of the function $f(z)$ equals 2, and
  similarly as in the proof of Lemma~\ref{orig:lem-9.3}, the type of
  $f$ equals $|\nu-\alpha|+\mu$. Obviously, the smallest type (namely,
  $\mu$) is achieved for $\alpha=\nu$. In particular, the function
  $\tilde \sigma(z)$ is of order $2$ and its type equals $\mu$.

  Conversely, suppose that the zero set of an entire function $f(z)$
  coincides with $\Lambda$ and that all zeroes of $f(z)$ are simple.
  Since $\tau_\Lambda=2$ the Hadamard theorem implies that
  $\varrho_f\ge 2$. Suppose that $\varrho_f=2$.  We can write $f(z)$
  in the form $f(z)=e^{g(z)}\sigma(z)$. By the Hadamard theorem,
  $g(z)=az^2+bz+c$. If $a=0$ then the type of $f(z)$ equals the type
  of $\sigma(z)$, if $a\ne0$ then the type of $f(z)$ equals the type
  of $\exp(az^2)\sigma(z)$. In the both cases the type of $f(z)$ is
  greater or equal $\mu$.
\end{proof}

\begin{rem}
  If $\Lambda$ is a quadratic or hexagonal lattice then $\nu=0$ and,
  consequently, $\tilde\sigma(z)=\sigma(z)$. Actually, in the former
  case we can suppose that $\omega_1>0$, $\omega_2=i\omega_1$. Then
  $\eta_1=\pi/\omega_1$, $\eta_2=-\pi{}i/\omega_1$ \cite[18.14.8 and
  18.14.10]{AS}, hence $\nu=0$. In the latter case we can suppose that
  $\omega_1=ke^{-i\pi/3}$, $\omega_2=ke^{i\pi/3}$, with $k>0$.  Then
  $$
  \eta_1=\frac{2\pi e^{i\pi/3}}{\sqrt{3}(\omega_1+\omega_2)}\,,
  \quad
  \eta_2=\frac{2\pi e^{-i\pi/3}}{\sqrt{3}(\omega_1+\omega_2)}\,
  $$
  \cite[18.13.16 and 18.13.19]{AS}. In this case, too, $\nu=0$.
\end{rem}

\begin{rem}
  There exist lattices for which $\nu\ne 0$ and, consequently,
  $\tilde\sigma(z)\ne\sigma(z)$. It suffices to consider a lattice
  with $\eta_2=0$ (such a lattice exists, see \cite[18.3.10]{AS}).
  Then, by the Lagrange formula, $|\eta_1\omega_2|=2\pi$ and hence
  $|\nu|=\pi/2S$. This means that the type of the $\sigma$-function
  for such a lattice equals $\pi/S$. Since $\nu$ depends on
  $(\omega_1,\omega_2)$ continuously any value of $|\nu|$ lying
  between $0$ and $\pi/2S$ is realized by a convenient lattice.
\end{rem}

\section*{Acknowledgements}

V.A.G. was supported by the Grants of RFBR (no. 02-01-00804), DFG--RAS
(no. 436 RUS 113/572/0-2) and INTAS (no. 00-257). P.\v{S}. gratefully
acknowledges support of the Ministry of Education of Czech Republic
under the research plan MSM210000018.

\end{document}